\documentclass[sigconf,natbib=false,timestamp=false]{acmart}
\pdfoutput=1

\RequirePackage[l2tabu, orthodox]{nag}

\usepackage[T1]{fontenc}

\usepackage{csvsimple}

\PassOptionsToPackage{lowtilde}{url}

\usepackage{paralist}

\usepackage{url}

\usepackage{xcolor}

\usepackage{multirow}

\usepackage{booktabs}

\usepackage{csquotes}

\usepackage{tabularx}

\usepackage{graphicx}
\usepackage{color}

\usepackage[ruled,lined,linesnumbered]{algorithm2e}
\usepackage[noend]{algorithmic}

\usepackage{array}

\usepackage{hhline}

\usepackage{verbatim}

\usepackage[export]{adjustbox}
\usepackage{listings}

\usepackage{mathtools}

\usepackage{xspace}

\usepackage{todonotes}
\newif\ifdraft\drafttrue

\usepackage{framed}

\usepackage{siunitx}

\newcommand{\head}[1]{\par\noindent\textbf{#1:}\space}

\newif\ifdoubleblind\doubleblindtrue
\newcommand{\dblblind}[2]{\ifdoubleblind#1\else#2\fi}

\usepackage[abbreviate=true, dateabbrev=true, isbn=false, doi=false, urldate=comp, url=false, maxbibnames=9, backref=false, backend=biber, bibencoding=utf8, style=ACM-Reference-Format, language=american, giveninits=true, sortcites=true, sorting=none]{biblatex}
\AtEveryBibitem{\clearlist{location}} %
\AtEveryBibitem{\clearlist{language}} %
\AtEveryBibitem{\clearlist{month}} %

\usepackage[type={CC}, modifier={by}, version={4.0}]{doclicense}

\doubleblindfalse
\newcommand{\CiscoNorway}[1]{\dblblind{#1}{Cisco Norway}} %

\newcommand{\Ncf}{\ensuremath{\text{N}_\text{CF}}} %
\newcommand{\Nuf}{\ensuremath{\text{N}_\text{UF}}} %
\newcommand{\Ncs}{\ensuremath{\text{N}_\text{CS}}} %
\newcommand{\Nus}{\ensuremath{\text{N}_\text{US}}} %

\newcommand{\cef}{\Ncf{}} %
\newcommand{\cnf}{\Nuf{}} %
\newcommand{\cep}{\Ncs{}} %
\newcommand{\cnp}{\Nus{}} %

\newcommand{\Aef}{\Ncf{}} %
\newcommand{\Anf}{\Nuf{}} %
\newcommand{\Aep}{\Ncs{}} %
\newcommand{\Anp}{\Nus{}} %

\newlength{\floatcorrection}
\copyrightyear{2020} 
\acmYear{2020} 
\setcopyright{rightsretained} 
\acmConference[ESEM '20]{ESEM '20: ACM / IEEE International Symposium on Empirical Software Engineering and Measurement (ESEM)}{October 8--9, 2020}{Bari, Italy}
\acmBooktitle{ESEM '20: ACM / IEEE International Symposium on Empirical Software Engineering and Measurement (ESEM) (ESEM '20), October 8--9, 2020, Bari, Italy}\acmDOI{10.1145/3382494.3410684}
\acmISBN{978-1-4503-7580-1/20/10}

\makeatletter
\def\@copyrightpermission{\doclicenseImage[imagewidth=2cm]\hspace*{1mm}\raisebox{3mm}{\parbox{6.3cm}{This work is licensed under a \doclicenseLongNameRef{} (\doclicenseNameRef) license.}}\vspace{1ex}} 
\makeatother

\begin{document}

\title{Spectrum-Based Log Diagnosis}

\author{Carl Martin Rosenberg}
\affiliation{%
  \institution{Simula Research Laboratory}
  \city{Oslo, Norway}
}
\email{cmr@simula.no}

\author{Leon Moonen}
\affiliation{%
  \institution{Simula Research Laboratory}
  \city{Oslo, Norway}
}
\email{leon.moonen@computer.org}

\ccsdesc[500]{Software and its engineering~Software development process management}
\ccsdesc[500]{Software and its engineering~Software defect analysis}

\begin{abstract}

\head{Background}
Continuous Engineering practices are increasingly adopted in modern software development.
However, a frequently reported need is for more effective methods to analyze the 
massive amounts of data resulting from the numerous build and test runs.

\head{Aims}
We present and evaluate Spectrum-Based Log Diagnosis (SBLD),
a method to help developers quickly diagnose problems found in complex integration and deployment runs.
Inspired by Spectrum-Based Fault Localization,
SBLD leverages the differences in event occurrences between logs for failing and passing runs, 
to highlight events that are stronger associated with failing runs. 

\head{Method}
Using data provided by \CiscoNorway{our industrial partner}, 
we empirically investigate the following questions: 
(i) How well does SBLD reduce the \emph{effort needed} to identify all \emph{failure-relevant events} in the log for a failing run?
(ii) How is the \emph{performance} of SBLD affected by \emph{available data}? 
(iii) How does SBLD compare to searching for \emph{simple textual patterns} that often occur in failure-relevant events?
We answer (i) and (ii) using summary statistics and heatmap visualizations, 
and for (iii) we compare three configurations of SBLD (with resp. minimum, median and maximum data) 
against a textual search using Wilcoxon signed-rank tests and the Vargha-Delaney measure of stochastic superiority.

\head{Results}
Our evaluation shows that 
(i) SBLD achieves a significant effort reduction for the dataset used,
(ii) SBLD benefits from additional logs for passing runs in general, 
and it benefits from additional logs for failing runs when there is a proportional amount of logs for passing runs in the data.
Finally, (iii) SBLD and textual search are roughly equally effective at effort-reduction, 
while textual search has slightly better recall. 
We investigate the cause, and discuss how it is due to characteristics of a specific part of our data.

\head{Conclusions}
We conclude that SBLD shows promise as a method for diagnosing failing runs, 
that its performance is positively affected by additional data, 
but that it does not outperform textual search on the dataset considered. 
Future work includes investigating SBLD's generalizability on additional datasets.

\end{abstract}

\keywords{%
continuous engineering,
failure diagnosis,
log analysis,
log mining.
}

\maketitle

\section{Introduction}

Continuous Engineering (CE) practices, 
such as Continuous Integration (CI) and Continuous Deployment (CD), 
are gaining prominence in software engineering, 
as they help streamline and optimize the way software is built, tested and shipped. 
The most salient advantage of CE is the tighter feedback loops: 
CE practices help developers test and build their software more, 
and makes software releases less brittle by enabling more incremental releases.

Nevertheless, a frequently reported barrier for success is the need to effectively analyze
the data that results from the numerous build and test
runs~\cite{Laukkanen2017,Hilton2017,Shahin2017,Debbiche2014,Olsson2012}.
One evident example of this is the handling and
analysis of results from complex end-to-end integration tests 
which we focus on in this paper: 
CE practices make it easier to run such end-to-end tests, 
which include system integration and deployment to production hardware, 
and they are critical for ensuring the quality of the end product. 
However, since these end-to-end tests by their nature can fail for multiple
reasons, not least in the sense that new product code can make the tests
fail in new ways, it is critical to rapidly diagnose these failures.

In this paper we concern ourselves with how to rapidly analyze a set
of logs resulting from complex CE tasks\footnote{~For simplicity, and without loss of generality, 
we will refer to these CE tasks as ``integration tests'' or ``tests'' throughout the paper, 
though we acknowledge that they include more than just testing, 
such as building the system and deploying it on hardware in a test or staging environment, 
and failures can occur in any of these phases. 
The proposed approach aims to cover all these situations, 
and is evaluated on real-life logs capturing everything from building the system, 
to deploying it on production hardware, 
and running complex integration and interaction scenarios.} 
where the overall outcome of the task (i.e. 'fail' or 'pass') is known, 
but where analysts must consult the resulting logs to fully diagnose why the failures occurred. 
Since these logs can get large and unwieldy, we
develop a tool that automatically suggests which segments in the logs
are most likely relevant for troubleshooting purposes. 
Our method gives each event in the log an interestingness score based
on the overall event frequencies in the test result set: The log
events are in turn clustered based on these scores, and the event
clusters are presented to the user in decreasing order of overall
interestingness. The goal is to enable users to find all relevant
diagnostic information in the first presented event cluster, while having the
option of retrieving additional clusters if needed. An
additional benefit of our method is that the extracted events can help
identify commonly occurring patterns that are symptomatic for specific
errors. Future logs that exhibit the same characteristics can then be
automatically classified as having symptoms of that error.

\head{Contributions} We present Spectrum-Based Log Diagnosis (SBLD), a method for helping developers quickly find the
most relevant segments of a log. Using data from \CiscoNorway{an
industrial partner}, we empirically evaluate SBLD by investigating the following
three questions:
(i) How well does SBLD reduce the \emph{effort needed} to identify all \emph{failure-relevant events} in the log for a failing run?
(ii) How is the \emph{performance} of SBLD affected by \emph{available data}?
(iii) How does SBLD compare to searching for \emph{simple textual patterns} that often occur in failure-relevant events?

\head{Overview} 
The rest of the paper is structured as follows: Section~\ref{sec:approach}
explains SBLD and the methodology underlying its event ranking
procedures. Sections~\ref{sec:rqs} and~\ref{sec:expdesign} motivates our research questions 
and empirical design. We report and discuss our results in
Section~\ref{sec:resdiscuss}. Section~\ref{sec:relwork} surveys related work,
and we discuss threats to validity in Section~\ref{sec:ttv} before concluding 
in Section~\ref{sec:conclusion}.

\section{Approach}
\label{sec:approach}

\begin{figure}[b]
        \includegraphics[width=0.99\columnwidth]{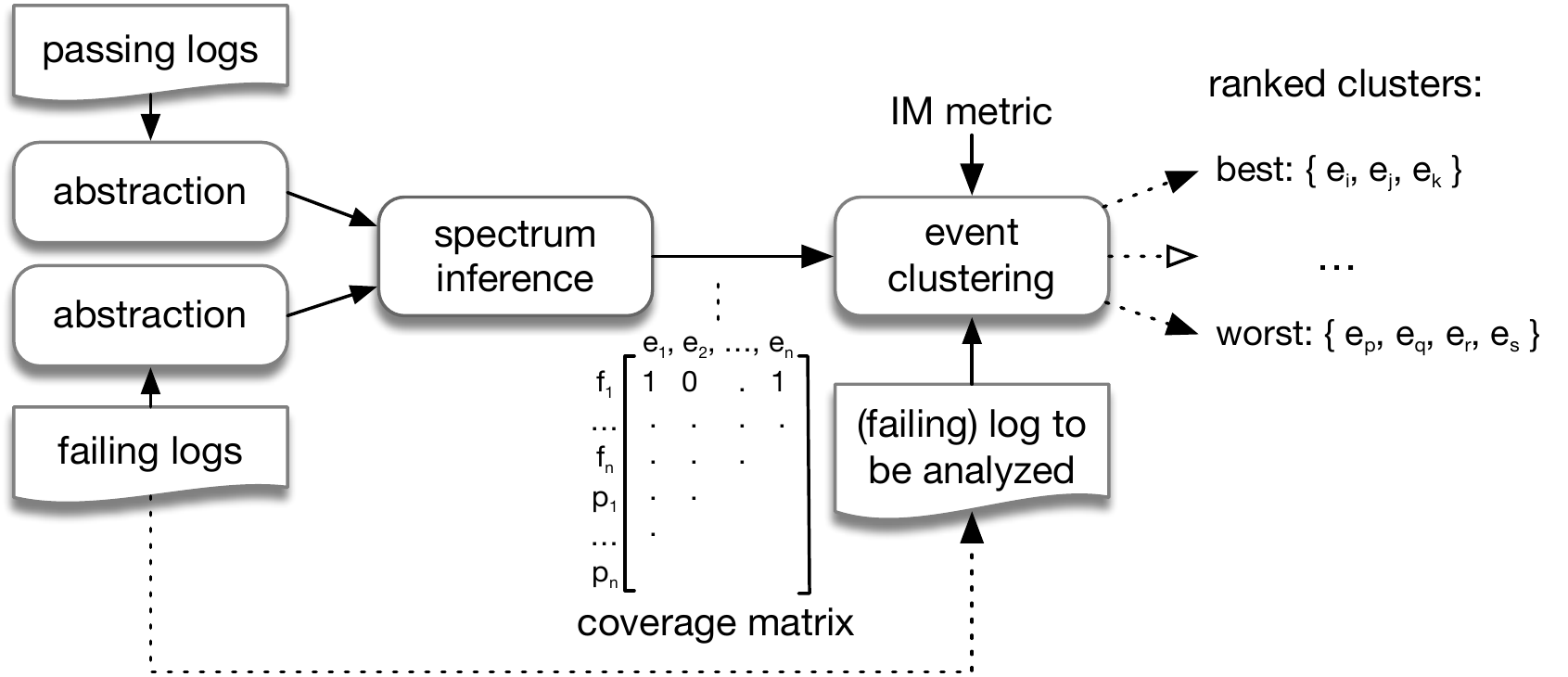}
        	\vspace*{-2ex}
        \caption{A visual overview of our approach.}
        \label{fig:approach}
\end{figure}

SBLD takes a set of log files from test failures, a set of log files from test successes, and a singular log file from a test failure called the \emph{target log} that the user wants analyzed and produces a list of segments  from the target log file that are likely relevant for understanding why the corresponding test run failed.

In the following we explain the workings of SBLD in a stepwise
manner. At each step, we present the technical background needed to
understand how SBLD accomplishes its task. A visual overview of SBLD is
shown in Figure \ref{fig:approach}.

\head{Prerequisites}
First of all, SBLD requires access to a set of log files from failing test runs and a set of log files from successful test runs.
For brevity, we will refer to log files from failing test runs as 'failing logs', 
and log files from successful test runs as 'passing logs'.%
\footnote{~Note that we explicitly assume that the outcome of each run is known; 
  This work is not concerned with determining whether the run was a failure or a success, 
  but rather with helping identify why the failing runs failed.} 
We also require a programmatic way of segmenting each log file
into individually meaningful components. For the dataset used in this
paper these components are \emph{events} in the form of blocks of text
preceded by a date and a time-stamp in a predictable format. Lastly,
we require that run-time specific information such as timestamps,
dynamically generated IP addresses, check-sums and so on are removed
from the logs and replaced with standardized text. We refer to the process of
enforcing these requirements and delineating the log into events as
the \emph{abstraction} step. This enables SBLD to treat events
like ``2019-04-05 19:19:22.441 CEST: Alice calls Bob'' and ``2019-04-07
13:12:11.337 CEST: Alice calls Bob'' as two instances of the same
generic event "Alice calls Bob". The appropriate degree of abstraction
and how to meaningfully delineate a log will be context-dependent
and thus we require the user to perform these steps before using SBLD. 
In the current paper we use an abstraction mechanism
and dataset generously provided by \CiscoNorway{our industrial partner}.

\renewcommand{\Ncf}{\ensuremath{\text{N}_\text{FI}}} %
\renewcommand{\Nuf}{\ensuremath{\text{N}_\text{FE}}} %
\renewcommand{\Ncs}{\ensuremath{\text{N}_\text{PI}}} %
\renewcommand{\Nus}{\ensuremath{\text{N}_\text{PE}}} %

\head{Computing coverage and event relevance} SBLD requires an assumption about what makes an event \emph{relevant}
and a method for computing this relevance. Our method takes inspiration
from Spectrum-Based Fault Localization (SBFL) in which the suspiciousness 
or fault-proneness of a program statement is treated as a function of 
the number of times the statement was activated in a failing test case, 
combined with the number of times it is skipped in a passing test case~\cite{Jones2002,Abreu2007,Abreu2009}. 
The four primitives that need to be computed are shown on the right-hand side in Table~\ref{table:measures}. 
We treat each abstracted event as a statement and study their occurrences
in the logs like Fault Localization tracks the activation of statements in test cases. 
We compute the analysis primitives by devising a binary
\emph{coverage matrix} whose columns represent every unique event
observed in the set of failing and successful logs while each row $r$
represents a log and tracks whether the event at column $c$ occurred in
log $r$ (1), or not (0), as shown in Figure~\ref{fig:approach}.

By computing these primitives, we can rank each event by using an
\emph{interestingness measure} (also referred to as ranking
metric, heuristic, or similarity coefficient~\cite{Wong2016}). 
The choice of interestingness measure
is ultimately left to the user, as these are context dependent and 
there is no generally optimal choice of interestingness measure~\cite{Yoo2014}. 
In this paper we consider a
selection of nine interestingness measures prominent in the literature
and a simple metric that emphasizes the events that exclusively occur
in failing logs in the spirit of the \emph{union model} discussed
by Renieres et al.~\cite{renieres2003:fault}. We
report on the median performance of these interestingness measures with the intention of providing a
representative, yet unbiased, result. The ten measures considered are
precisely defined in Table~\ref{table:measures}.

\begin{table*}
\centering
\begin{tabular}{c@{\hspace{10mm}}c}
{\renewcommand{\arraystretch}{1.7} %
\begin{tabular}{lc}
\toprule
measure	 & formula \\\midrule

Tarantula \cite{Jones2001,Jones2002} & %

\( \frac{ \frac{ \cef{} }{ \cef{} + \cnf{} } }{ \frac{ \cef{} }{ \cef{} + \cnf{} } + \frac{ \cep{} }{ \cep{} + \cnp{} } } \) 

\\

Jaccard \cite{Jaccard1912,Chen2002} & %

\( \frac{ \Ncf }{ \Ncf + \Nuf + \Ncs } \)

\\

Ochiai \cite{Ochiai1957,Abreu2006} & %

\( \frac{ \Ncf }{ \sqrt{ ( \cef + \cnf ) \times ( \cef + \cep ) } } \)

\\

Ochiai2 \cite{Ochiai1957, Naish2011} & %

\( \frac{ \Aef \times \Anp }{ \sqrt{ ( \Aef + \Aep ) \times ( \Anf + \Anp ) \times ( \Aef + \Anf) \times ( \Aep + \Anp ) } } \)

\\

Zoltar \cite{Gonzalez2007} & %

\( \frac{ \Ncf }{ \Ncf + \Nuf + \Ncs + \frac { 10000 \times \Nuf \times \Ncs }{ \Ncf } } \)

\\

D$^\star$ \cite{Wong2014} (we use $\star = 2$) & %

\( \frac{ (\cef)^\star }{ \cnf + \cep } \)

\\

O$^p$ \cite{Naish2011} & %

\( \Aef - \frac{ \Aep }{ \Aep + \Anp + 1} \)

\\

Wong3 \cite{Wong2007,Wong2010} &

\( \Aef - h, \text{where~} h = \left\{ 
\scalebox{.8}{\(\renewcommand{\arraystretch}{1} %
\begin{array}{@{}ll@{}}
\Aep & \text{if~} \Aep \leq 2 \\
2 + 0.1(\Aep - 2) & \text{if~} 2 < \Aep \leq 10 \\
2.8 + 0.001(\Aep - 10) & \text{if~} \Aep > 10 \\
\end{array}\)}
\right. \)

\\

Kulczynski2 \cite{Kulczynski1927,Naish2011} & %

\( \frac{ 1 }{ 2 } \times ( \frac{ \Aef }{ \Aef + \Anf } + \frac{ \Aef }{ \Aef + \Aep } ) \)

\\

Failed only & %

\( \left\{\scalebox{.8}{\(\renewcommand{\arraystretch}{1} %
\begin{array}{@{}ll@{}}
1 & \text{if~} \Ncs = 0 \\
0 & \text{otherwise~}  \\
\end{array}\)}
\right. \)

\\
\bottomrule

\end{tabular}} &
\begin{tabular}{lp{2.99cm}}
\toprule
\multicolumn{2}{l}{notation used} \\\midrule
\Ncf & number of \emph{failing} logs \\ & that \emph{include} the event \\
\Nuf & number of \emph{failing} logs \\ & that \emph{exclude} the event \\
\Ncs & number of \emph{passing} logs \\ & that \emph{include} the event \\
\Nus & number of \emph{passing} logs \\ & that \emph{exclude} the event \\
\bottomrule
\end{tabular}
\end{tabular}\vspace*{1ex}
\caption{\label{table:measures}The 10 interestingness measures under consideration in this paper.}
\vspace*{-3ex}
\end{table*}

\head{Analyzing a target log file} Using our database of event scores,
we first identify the events occurring in the target log file and the
interestingness scores associated with these events. Then, we group
similarly scored events together using a clustering algorithm. Finally,
we present the best performing cluster of events to the end user. The
clustering step helps us make a meaningful selection of events rather
than setting an often arbitrary window selection size. Among other
things, it prevents two identically scored events from falling at
opposite sides of the selection threshold. If the user suspects that
the best performing cluster did not report all relevant events, she can
inspect additional event clusters in order of decreasing
aggregate interestingness score.  To perform the clustering step we use Hierarchical Agglomerative
Clustering (HAC) with Complete linkage~\cite{manning2008introduction}, where
sub-clusters are merged until the maximal distance between members of
each candidate cluster exceeds some specified threshold. In SBLD,
this threshold is the uncorrected sample standard deviation of the event
scores for the events being clustered.\footnote{~Specifically, 
we use the \texttt{numpy.std} procedure from the SciPy framework~\cite{2020SciPy-NMeth},
in which the uncorrected sample standard deviation is given by
$ \sqrt{\frac{1}{N} \sum_{i=1}^{N}\lvert x_{i} - \bar{x} \rvert^2} $ where
$\bar{x}$ is the sample mean of the interestingness scores obtained for the
events in the log being analyzed and $N$ is the number of events in the log.}  
This ensures that the ``interestingness-distance'' between two events 
in a cluster never exceeds the uncorrected sample standard deviation observed in the set.

\section{Research Questions}
\label{sec:rqs}

The goal of this paper is to present SBLD and help practitioners make
an informed decision whether SBLD meets their needs. To this end, we have identified
three research questions that encompass several concerns practitioners
are likely to have and that also are of interested to the research community at
large:
\begin{enumerate}[\bfseries RQ1]

\item How well does SBLD reduce the effort needed to identify all
        known-to-be relevant events ("does it work?") ?

\item How is the efficacy of SBLD impacted by increased evidence in the form of
        additional failing and passing logs ("how much data do we need before
                running the analysis?") ?

\item How does SBLD perform compared to a strategy based on searching for
        common textual patterns with a tool like \texttt{grep} ("is it better than doing the obvious thing?") ?
\end{enumerate}
RQ1 looks at the aggregated performance of SBLD to assess its viability.
With RQ2 we assess how sensitive the performance is to the amount of
available data: How many logs should you have before you can expect the
analysis to yield good results? Is more data unequivocally a good thing?
What type of log is more informative: A passing log or a failing log?
Finally, we compare SBLD's performance to a more traditional method for
finding relevant segments in logs: Using a textual search for strings 
one expects to occur near informative segments, like
"failure" and "error". The next section details the dataset used, our
chosen quality measures for assessment and our methodology for answering
each research question.

\section{Experimental Design}
\label{sec:expdesign}

\begin{table}
\centering
\caption{The key per-test attributes of our dataset. Two events are considered
        distinct if they are treated as separate events after the abstraction
        step. A "mixed" event is an event that occurs in logs of both failing and
        passing runs.}
\vspace*{-1ex}
\label{table:descriptive}
\renewcommand{\tabcolsep}{0.11cm}\small
\begin{tabular}{rcrrrrrr}
\toprule
     &           & \# fail   & \# pass   & distinct & fail-only  & mixed   & pass-only  \\
test & signature & logs      & logs      &  events  & events     & events  & events \\
\midrule
   1 & C  &     24 &     100 &           36391 &            21870 &          207 &               14314 \\
   2 & E  &     11 &      25 &             380 &               79 &          100 &                 201 \\
   3 & E  &     11 &      25 &             679 &              174 &           43 &                 462 \\
   4 & E  &      4 &      25 &             227 &               49 &           39 &                 139 \\
   5 & C  &      2 &     100 &           33420 &             2034 &           82 &               31304 \\
   6 & C  &     19 &     100 &           49155 &            15684 &          893 &               32578 \\
   7 & C  &     21 &     100 &           37316 &            17881 &          154 &               19281 \\
   8 & C  &      4 &     100 &           26614 &             3976 &           67 &               22571 \\
   9 & C  &     21 &     100 &           36828 &            19240 &          228 &               17360 \\
  10 & C  &     22 &     100 &          110479 &            19134 &         1135 &               90210 \\
  11 & E  &      5 &      25 &             586 &               95 &           47 &                 444 \\
  12 & E  &      7 &      25 &             532 &               66 &           18 &                 448 \\
  13 & C  &      2 &     100 &           15351 &             2048 &          232 &               13071 \\
  14 & C  &      3 &     100 &           16318 &             2991 &          237 &               13090 \\
  15 & C  &     26 &     100 &           60362 &            20964 &         1395 &               38003 \\
  16 & C  &     12 &     100 &            2206 &              159 &          112 &                1935 \\
  17 & E  &      8 &      25 &             271 &               58 &           98 &                 115 \\
  18 & A  &     23 &      75 &            3209 &              570 &          156 &                2483 \\
  19 & C  &     13 &     100 &           36268 &            13544 &          411 &               22313 \\
  20 & B  &      3 &      19 &             688 &               69 &           31 &                 588 \\
  21 & B  &     22 &      25 &             540 &              187 &           94 &                 259 \\
  22 & E  &      1 &      25 &             276 &               11 &           13 &                 252 \\
  23 & C  &     13 &     100 &           28395 &            13629 &          114 &               14652 \\
  24 & E  &      7 &      26 &             655 &              117 &           56 &                 482 \\
  25 & C  &     21 &     100 &           44693 &            18461 &          543 &               25689 \\
  26 & C  &     21 &     100 &           42259 &            19434 &          408 &               22417 \\
  27 & C  &     21 &     100 &           44229 &            18115 &          396 &               25718 \\
  28 & C  &     20 &     100 &           43862 &            16922 &          642 &               26298 \\
  29 & C  &     28 &     100 &           54003 &            24216 &         1226 &               28561 \\
  30 & C  &     31 &     100 &           53482 &            26997 &         1063 &               25422 \\
  31 & C  &     27 &     100 &           53092 &            23283 &          463 &               29346 \\
  32 & C  &     21 &     100 &           55195 &            19817 &          768 &               34610 \\
  33 & E  &      9 &      25 &             291 &               70 &           30 &                 191 \\
  34 & D  &      2 &      13 &             697 &               76 &           92 &                 529 \\
  35 & E  &      9 &      25 &             479 &              141 &           47 &                 291 \\
  36 & E  &     10 &      75 &            1026 &              137 &           68 &                 821 \\
  37 & E  &      7 &      25 &            7165 &             1804 &           94 &                5267 \\
  38 & E  &      4 &      25 &             647 &               67 &           49 &                 531 \\
  39 & G  &     47 &     333 &            3350 &              428 &          144 &                2778 \\
  40 & G  &     26 &     333 &            3599 &              240 &          157 &                3202 \\
  41 & G  &     26 &     332 &            4918 &              239 &          145 &                4534 \\
  42 & C  &     17 &     100 &           30411 &            14844 &          348 &               15219 \\
  43 & F  &    267 &     477 &           10002 &             3204 &         1519 &                5279 \\
  44 & C  &      9 &     100 &           29906 &             8260 &          274 &               21372 \\
  45 & E  &      3 &      25 &             380 &               44 &           43 &                 293 \\
\bottomrule
\end{tabular}
\vspace*{-2ex}
\end{table}

\begin{table}
\centering
\caption{Ground-truth signatures and their occurrences in distinct events.}
\label{table:signature}
\vspace*{-1ex}
\small
\begin{tabular}{ccrrrc}
\toprule
          &   sub-  & fail-only & pass-only & fail \& & failure \\
signature & pattern & events    & events    & pass    & strings* \\
\midrule
        A &       1 &                1 &                0 &            0 &                                  yes \\
        A &       2 &                2 &                0 &            0 &                                   no \\
        B &       1 &                2 &                0 &            0 &                                  yes \\
        C &       1 &               21 &                0 &            0 &                                  yes \\
        C &       2 &               21 &                0 &            0 &                                  yes \\
        D &       1 &                4 &                0 &            0 &                                  yes \\
 \textbf{D$^{\#}$} & \textbf{2} &               69 &              267 &          115 &                                   no \\
 \textbf{D$^{\#}$} & \textbf{3} &                2 &               10 &           13 &                                   no \\
 \textbf{E$^{\#}$} & \textbf{1} &               24 &              239 &          171 &                                   no \\
        E &       1 &                1 &                0 &            0 &                                   no \\
        E &       2 &                9 &                0 &            0 &                                   no \\
        E &       3 &                9 &                0 &            0 &                                  yes \\
        E &       4 &               23 &                0 &            0 &                                  yes \\
        F &       1 &               19 &                0 &            0 &                                  yes \\
        F &       2 &               19 &                0 &            0 &                                   no \\
        F &       3 &               19 &                0 &            0 &                                  yes \\
        F &       4 &               14 &                0 &            0 &                                  yes \\
        G &       1 &                2 &                0 &            0 &                                  yes \\
        G &       2 &                1 &                0 &            0 &                                   no \\
        G &       3 &                1 &                0 &            0 &                                   no \\
\bottomrule
\multicolumn{6}{l}{* signature contains the lexical patterns 'error', 'fault' or 'fail*'}\\
\multicolumn{6}{l}{$^{\#}$ sub-patterns that were removed to ensure a clean ground truth}
\end{tabular}
\vspace*{-3ex}
\end{table}
 
\subsection{Dataset and ground truth}
\label{sec:dataset}

Our dataset provided by \CiscoNorway{our industrial partner} consists
of failing and passing log files from 45 different end-to-end integration
tests. In addition to the log text we also have data on when a given
log file was produced. Most test-sets span a time-period of 38 days, while
the largest set (test 43 in Table~\ref{table:descriptive}) spans 112
days. Each failing log is known to exemplify symptoms of one of seven
known errors, and \CiscoNorway{our industrial partner} has given us a
set of regular expressions that help determine which events are relevant
for a given known error. We refer to the set of regular expressions
that identify a known error as a \emph{signature} for that error. These
signatures help us construct a ground truth for our investigation.
Moreover, an important motivation for developing SBLD is to help create
signatures for novel problems: The events highlighted by SBLD should be
characteristic of the observed failure, and the textual contents of the
events can be used in new signature expressions.

Descriptive facts about our dataset is listed in
Table~\ref{table:descriptive} while Table~\ref{table:signature}
summarizes key insights about the signatures used.

Ideally, our ground truth should highlight exactly and \emph{only} the
log events that an end user would find relevant for troubleshooting
an error. However, the signatures used in this investigation were
designed to find sufficient evidence that the \emph{entire log} in
question belongs to a certain error class: the log might contain other
events that a human user would find equally relevant for diagnosing
a problem, but the signature in question might not encompass these
events. Nevertheless, the events that constitute sufficient evidence
for assigning the log to a given error class are presumably relevant
and should be presented as soon as possible to the end user. However,
if our method cannot differentiate between these signature events and
other events we cannot say anything certain about the relevance of
those other events. This fact is reflected in our choice of quality
measures, specifically in how we assess the precision of the approach. This
is explained in detail in the next section.

When producing the ground truth, we first ensured that a log would only be
associated with a signature if the entire log taken as a whole satisfied all
the sub-patterns of that signature. If so, we then determined which events
the patterns were matching on. These events constitute the known-to-be relevant
set of events for a given log.  However, we identified some problems with two of the provided
signatures that made them unsuitable for assessing SBLD. Signature \emph{E}
(see Table~\ref{table:signature}) had a sub-pattern that searched for a "starting test"-prefix that necessarily
matches on the first event in all logs due to the structure of the logs.
Similarly, signature \emph{D} contained two sub-patterns that necessarily
match all logs in the set--in this case by searching for whether the test
was run on a given machine, which was true for all logs for the corresponding
test. We therefore elected to remove these sub-patterns from the signatures
before conducting the analysis.

\subsection{Quality Measures}

As a measure of how well SBLD reports all known-to-be relevant log
events, we measure \emph{recall in best cluster}, which we for brevity refer to
as simply \emph{recall}. 
This is an adaption of the classic recall measure used in information retrieval,
which tracks the proportion of all relevant events that were retrieved
by the system~\cite{manning2008introduction}. 
As our method presents events to the user in a series of ranked clusters, 
we ideally want all known-to-be relevant events to appear in the highest ranked cluster. 
We therefore track the overall recall obtained as if the first cluster were the only events retrieved.
Note, however, that SBLD ranks all clusters, and a user can retrieve additional clusters if desired. 
We explore whether this could improve SBLD's performance on a
specific problematic test-set in Section~\ref{sec:testfourtythree}.

It is trivial to obtain a perfect recall by simply retrieving all events
in the log, but such a method would obviously be of little help to a user
who wants to reduce the effort needed to diagnose failures.
We therefore also track the \emph{effort reduction} (ER), defined as
\[ \text{ER} = 1 - \frac{\text{number of events in first cluster}}{\text{number of events in log}} \]

Much like effective information retrieval systems aim for high recall and
precision, we want our method to score a perfect recall while obtaining the
highest effort reduction possible. 

\subsection{Recording the impact of added data}

To study the impact of added data on SBLD's performance, we need to measure how
SBLD's performance on a target log $t$ is affected by adding an extra
failing log $f$ or a passing log $p$. There are several strategies
for accomplishing this. One way is to try all combinations in the
dataset i.e.\ compute the performance on any $t$ using any choice of
failing and passing logs to produce the interestingness scores. This
approach does not account for the fact that the logs in the data are
produced at different points in time and is also extremely expensive
computationally. We opted instead to order the logs chronologically and
simulate a step-wise increase in data as time progresses, as shown in
Algorithm~\ref{alg:time}.

\begin{algorithm}[b]
\caption{Pseudo-code illustrating how we simulate a step-wise increase in data
        as time progresses and account for variability in choice of
        interestingness measure.}
\label{alg:time}
\begin{algorithmic}\small
\STATE $F$ is the set of failing logs for a given test
\STATE $P$ is the set of passing logs for a given test
\STATE $M$ is the set of interestingness measures considered
\STATE sort $F$ chronologically
\STATE sort $P$ chronologically
\FOR{$i=0$ to $i=\lvert F \rvert$}
        \FOR{$j=0$ to $j=\lvert P \rvert$}
                \STATE $f = F[:i]$ \COMMENT{get all elements in F up to and including position i}
                \STATE $p = P[:j]$
                \FORALL{$l$ in $f$}
                        \STATE initialize $er\_scores$ as an empty list
                        \STATE initialize $recall\_scores$ as an empty list
                        \FORALL{$m$ in $M$}
                                \STATE perform SBLD on $l$ using $m$ as measure \\ \hspace*{1.75cm} and $f$ and $p$ as spectrum data
                                \STATE append recorded effort reduction score to $er\_scores$
                                \STATE append recorded recall score to $recall\_scores$
                        \ENDFOR
                        \STATE record median of $er\_scores$
                        \STATE record median of $recall\_scores$
                \ENDFOR
        \ENDFOR
\ENDFOR
\end{algorithmic}
\end{algorithm}

\subsection{Variability in interestingness measures}
\label{sec:imvars}

As mentioned in Section~\ref{sec:approach}, SBLD requires a
choice of interestingness measure for scoring the events, 
which can have a considerable impact on SBLD's performance. 
Considering that the best choice of interestingness measure is context-dependent, 
there is no global optimum, 
it is up to the user to decide which interestingness metric best reflects their
notion of event relevance. 

Consequently, we want to empirically study SBLD in way
that captures the variability introduced by this decision. 
To this end, we record the median score obtained by performing SBLD for every possible choice of
interestingness measure from those listed in Table~\ref{table:measures}.
Algorithm~\ref{alg:time} demonstrates the procedure in pseudo-code.

\subsection{Comparing alternatives}
\label{sec:comps}

To answer RQ2 and RQ3, we use pairwise comparisons of
different configurations of SBLD with a method that searches for regular expressions. 
The alternatives are compared
on each individual failing log in the set in a paired fashion. An
important consequence of this is that the statistical comparisons have
no concept of which test the failing log belongs to, and thus the test
for which there is most data has the highest impact on the result of the
comparison.

The pairwise comparisons are conducted using paired Wilcoxon signed-rank
tests~\cite{wilcoxon1945} where the Pratt correction~\cite{Pratt1959}
is used to handle ties. We apply Holm's correction~\cite{Holm1979}
to the obtained p-values to account for the family-wise error
rate arising from multiple comparisons. We declare a comparison
\emph{statistically significant} if the Holm-adjusted p-value is below
$\alpha=0.05$. The Wilcoxon tests check the two-sided null hypothesis of
no difference between the alternatives. We report the Vargha-Delaney $A_{12}$ and
$A_{21}$~\cite{Vargha2000} measures of stochastic superiority to
indicate which alternative is the strongest. Conventionally, $A_{12}=0.56$ is
considered a small difference, $A_{12}=.64$ is considered a medium difference
and $A_{12}=.71$ or greater is considered large~\cite{Vargha2000}. Observe
also that $A_{21} = 1 - A_{12}$.

\begin{figure*}
        \includegraphics[width=0.8\textwidth]{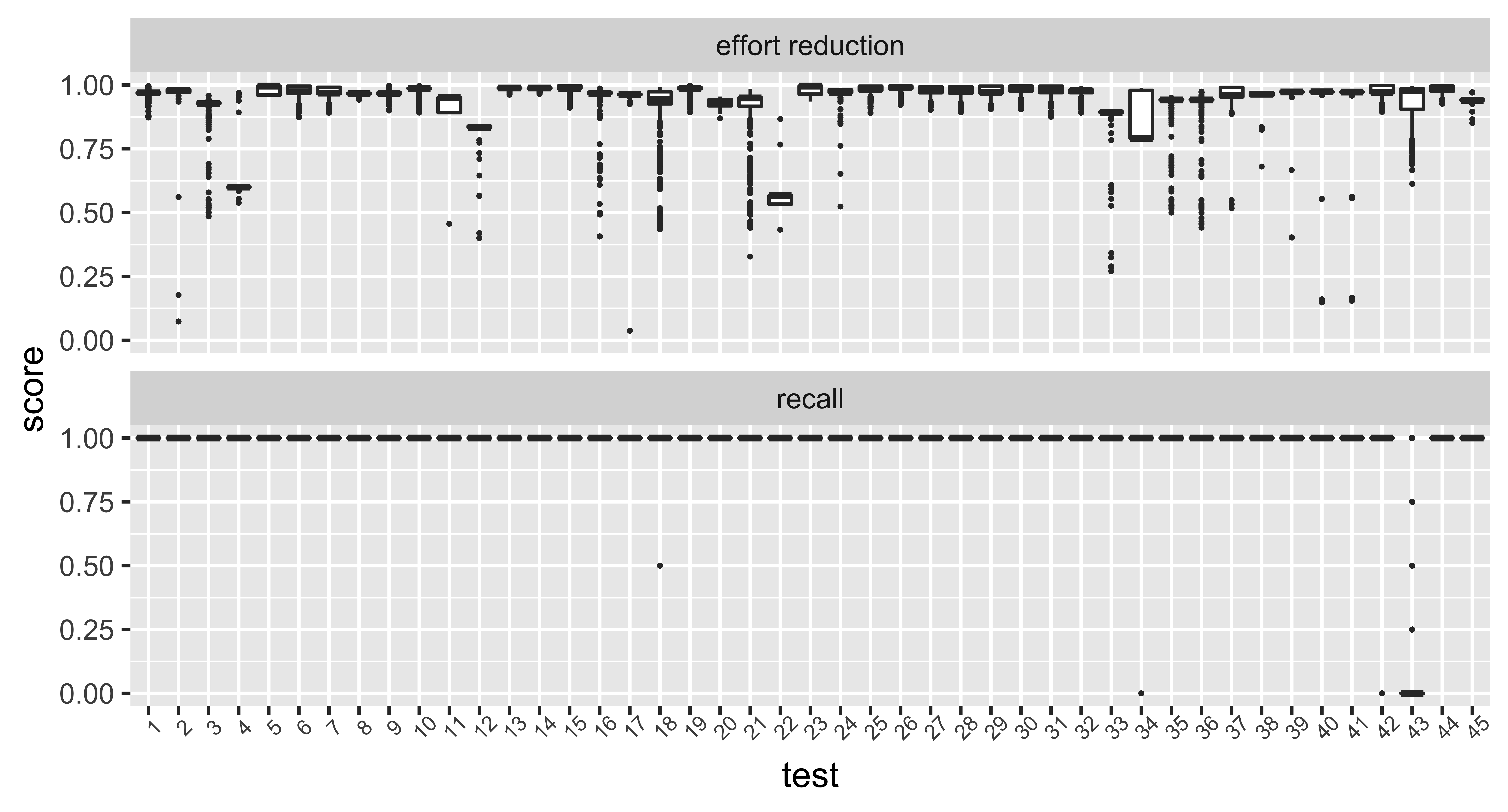}
        \caption{The overall performance of SBLD in terms of effort reduction
        and recall. On many tests, SBLD exhibited perfect recall for
        all observations in the inter-quartile range and thus the box collapses to a single line on the $1.0$ mark.\label{fig:rq1boxplot}}
\end{figure*}

\subsection{Analysis procedures}

We implement the SBLD approach in a prototype tool 
DAIM (Diagnosis and Analysis using Interestingness Measures), 
and use DAIM to empirically evaluate the idea.

\head{RQ1 - overall performance} We investigate the overall performance
of SBLD by analyzing a boxplot for each test in our dataset. Every individual
datum that forms the basis of the plot is the median performance of SBLD over
all choices of interestingness measures for a given set of failing and passing
logs subject to the chronological ordering scheme outlined above.

\head{RQ2 - impact of data} We analyze the impact of added data by
producing and evaluating heatmaps that show the obtained performance
as a function of the number of failing logs (y-axis) and number of
passing logs (x-axis). The color intensity of each tile in the heatmaps
is calculated by taking the median of the scores obtained for each
failing log analyzed with the given number of failing and passing logs
as data for the spectrum inference, wherein the score for each log is
the median over all the interestingness measures considered as outlined in
Section~\ref{sec:imvars}.

Furthermore, we compare three variant configurations
of SBLD that give an overall impression of the influence of added
data. The three configurations considered are \emph{minimal evidence},
\emph{median evidence} and \emph{maximal evidence}, where minimal
evidence uses only events from the log being analyzed and one additional
passing log, median evidence uses the median amount of respectively failing and
and passing logs available while maximal evidence uses
all available data for a given test. The comparisons are conducted with the
statistical scheme described above in Section~\ref{sec:comps}.

\head{RQ3 - SBLD versus pattern-based search} To compare SBLD
against a pattern-based search, we record the effort reduction and
recall obtained when only selecting events in the log that match on the
case-insensitive regular expression \texttt{"error|fault|fail*"}, where
the $*$ denotes a wildcard-operator and the $\lvert$ denotes logical
$OR$. This simulates the results that a user would obtain by using
a tool like \texttt{grep} to search for words like 'error' and 'failure'.
Sometimes the ground-truth signature expressions contain words from this
pattern, and we indicate this in Table~\ref{table:signature}. If so, the
regular expression-based method is guaranteed to retrieve the event.
Similarly to RQ2, we compare the three configurations of SBLD described
above (minimum, median and maximal evidence) against the pattern-based
search using the statistical described in Section~\ref{sec:comps}.

\section{Results and Discussion}
\label{sec:resdiscuss}

\begin{figure*}
\includegraphics[width=\textwidth]{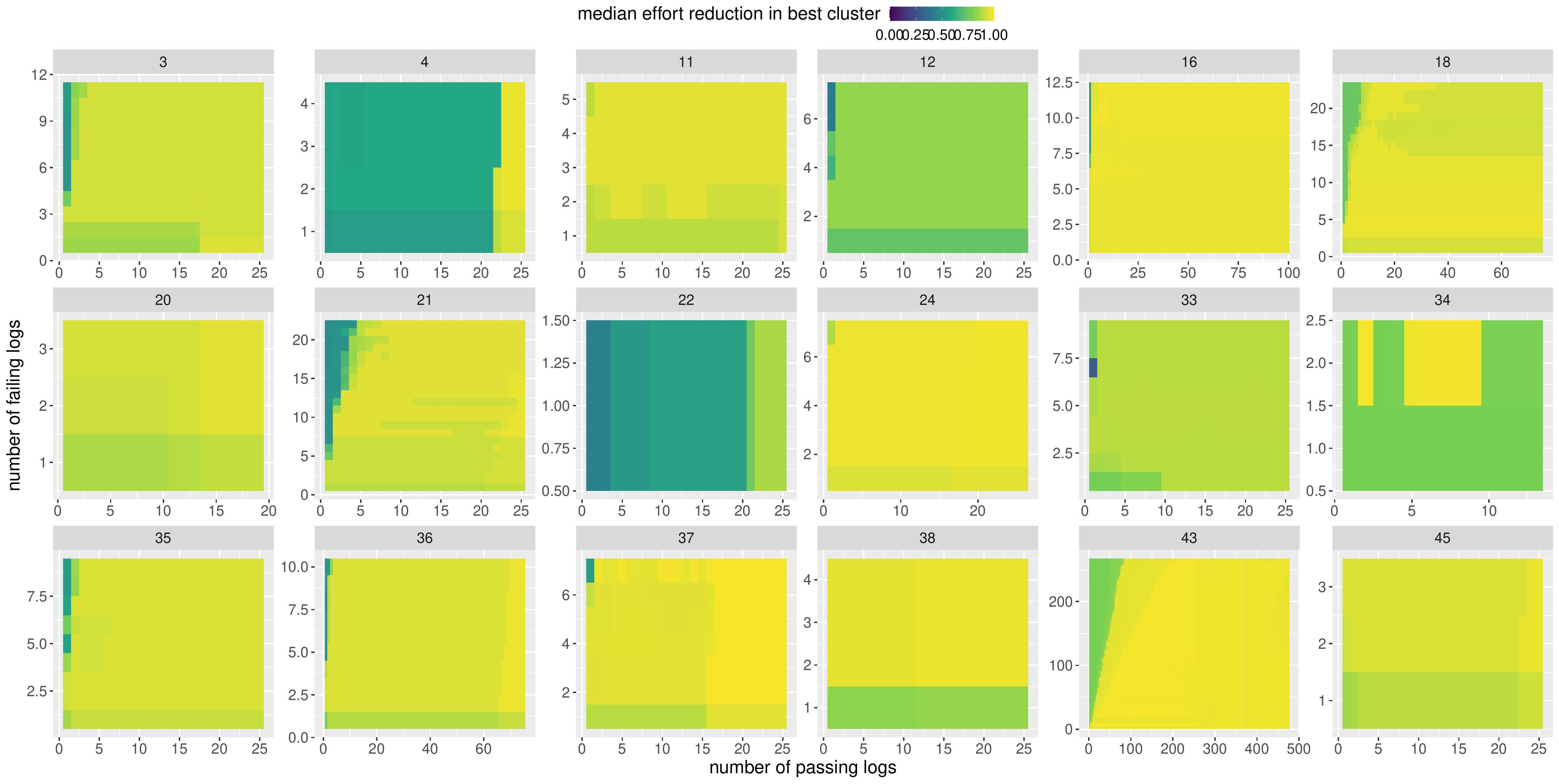}
        \caption{Effort reduction score obtained when SBLD is run on a given number of failing and passing logs. The tests not listed in this figure all obtained a lowest median effort reduction score of 90\% or greater and are thus not shown for space considerations. \label{fig:erheat}}
\vspace*{-2ex}
\end{figure*}

\begin{figure}
\includegraphics[width=\columnwidth]{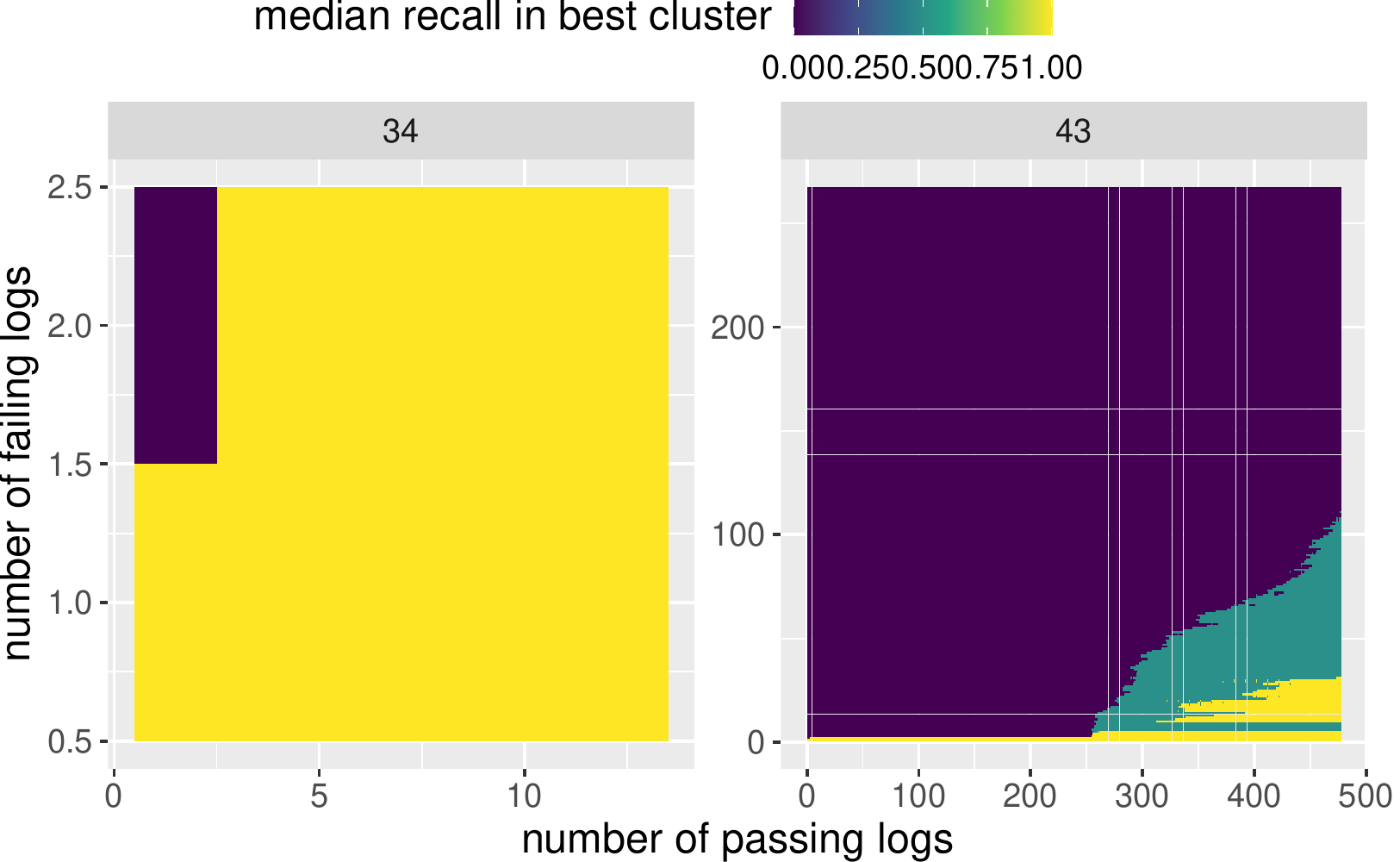}
        \caption{Recall score obtained when SBLD is run on a given number of failing and passing logs. For space
        considerations, we only show tests for which the minimum observed
        median recall was smaller than 1 (SBLD attained perfect median recall for all configurations in the other tests). \label{fig:recallheat}}
\vspace*{-3ex}
\end{figure}

This section gradually dissects Figure~\ref{fig:rq1boxplot}, showing a breakdown of SBLD's performance per test for both recall
and effort reduction, Figures \ref{fig:erheat} and \ref{fig:recallheat}, 
showing SBLD's performance as a function of the number of failing and passing
logs used, as well as Table~\ref{table:comparisons}, which shows the results
of the statistical comparisons we have performed.

\begin{table*}
\caption{Statistical comparisons performed in this investigation. The
bold p-values are those for which no statistically significant difference under $\alpha=0.05$
        could be established.}
\label{table:comparisons}
{\small%
\begin{tabular}{lllrrrr}
\toprule
 variant 1 & variant 2 &            quality measure &  Wilcoxon statistic &   $A_{12}$ &  $A_{21}$   & Holm-adjusted p-value\\
\midrule
 pattern-based search &  minimal  evidence &  effort reduction & 29568.5 & 0.777 & 0.223 &    $\ll$ 0.001 \\
 pattern-based search &  maximal  evidence &  effort reduction & 202413.0 & 0.506 & 0.494 &       \textbf{1.000} \\
 pattern-based search &   median evidence &  effort reduction & 170870.5 & 0.496 & 0.504 &    $\ll$ 0.001 \\
     minimal evidence &  maximal evidence &  effort reduction & 832.0 & 0.145 & 0.855 &    $\ll$ 0.001 \\
     minimal evidence &   median evidence &  effort reduction & 2666.0 & 0.125 & 0.875 &    $\ll$ 0.001 \\
     maximal evidence &   median evidence &  effort reduction & 164674.0 & 0.521 & 0.479 &       \textbf{1.000} \\
 pattern-based search &  minimal evidence &            recall & 57707.0 & 0.610 & 0.390 &    $\ll$ 0.001 \\
 pattern-based search &  maximal evidence &            recall & 67296.0 & 0.599 & 0.401 &    $\ll$ 0.001 \\
 pattern-based search &   median evidence &            recall & 58663.5 & 0.609 & 0.391 &    $\ll$ 0.001 \\
     minimal evidence &  maximal evidence &            recall & 867.5 & 0.481 & 0.519 &    $\ll$ 0.001 \\
     minimal evidence &   median evidence &            recall &             909.0 & 0.498 & 0.502 &       0.020 \\
     maximal evidence &   median evidence &            recall & 0.0 & 0.518 & 0.482 &    $\ll$ 0.001 \\
\bottomrule
\end{tabular}
}
\end{table*}

\subsection{RQ1: The overall performance of SBLD}

Figure~\ref{fig:rq1boxplot} suggests that SBLD's overall performance is strong,
since it obtains near-perfect recall while retaining a high degree of effort
reduction.  In terms of recall, SBLD obtains a perfect performance on all except
four tests: 18, 34, 42 and 43, with the lower quartile stationed at perfect recall for all tests
except 43 (which we discuss in detail in Section~\ref{sec:testfourtythree}).
For test 18, only 75 out of 20700 observations ($0.036\%$) obtained a recall score
of $0.5$ while the rest obtained a perfect score. On test 34 (the smallest in our
dataset), 4 out of 39 observations obtained a score of zero recall while the
others obtained perfect recall. 
For test 42, 700 out of 15300 ($0.4\%$) observations obtained a score of zero recall while the rest obtained perfect recall.
Hence with the exception of test 43 which is discussed later, 
SBLD obtains very strong recall scores overall with only a few outliers.

The performance is also strong in terms of effort reduction, albeit
more varied. To a certain extent this is expected since the attainable
effort reduction on any log will vary with the length of the log and
the number of ground-truth relevant events in the log. As can be seen
in Figure~\ref{fig:rq1boxplot}, most of the observations fall well
over the 75\% mark, with the exceptions being tests 4 and 22. For test
4, Figure~\ref{fig:erheat} suggests that one or more of the latest
passing logs helped SBLD refine the interestingness scores. A similar
but less pronounced effect seems to have happened for test 22. However,
as reported in Table~\ref{table:descriptive}, test 22 consists only of
\emph{one} failing log. Manual inspection reveals that the log consists
of 30 events, of which 11 are fail-only events. Without additional
failing logs, most interestingness measures will give a high score to
all events that are unique to that singular failing log, which is likely
to include many events that are not ground-truth relevant. Reporting 11
out of 30 events to the user yields a meager effort reduction of around
63\%. Nevertheless, the general trend is that SBLD retrieves a compact
set of events to the user which yields a high effort reduction score.

In summary, the overall performance shows that SBLD
retrieves the majority of all known-to-be-relevant events
in compact clusters, which dramatically reduces the analysis burden for the
end user. The major exception is Test 43, which we return to in
Section~\ref{sec:testfourtythree}.

\subsection{RQ2: On the impact of evidence}

The heatmaps suggest that the effort reduction is generally not
adversely affected by adding more \emph{passing logs}. If the
assumptions underlying our interestingness measures are correct,
this is to be expected: Each additional passing log either gives us
reason to devalue certain events that co-occur in failing and passing
logs or contain passing-only events that are deemed uninteresting.
Most interestingness measures highly value events that
exclusively occur in failing logs, and additional passing logs help
reduce the number of events that satisfy this criteria. However, since
our method bases itself on clustering similarly scored events it is
weak to \emph{ties} in interestingness scores. It is possible that
an additional passing log introduces ties where there previously was
none. This is likely to have an exaggerated effect in situations with
little data, where each additional log can have a dramatic impact on the
interestingness scores. This might explain the gradual dip in effort
reduction seen in Test 34, for which there are only two failing logs.

Adding more failing logs, on the other hand, draws a more nuanced
picture: When the number of failing logs (y-axis) is high relative
to the number of passing logs (x-axis), effort reduction seems to suffer.
Again, while most interestingness measures will prioritize events that
only occur in failing logs, this strategy only works if there is a
sufficient corpus of passing logs to weed out false positives. When
there are far fewer passing than failing logs, many events will be
unique to the failing logs even though they merely reflect a different
valid execution path that the test can take. This is especially true for
complex integration tests like the ones in our dataset, which might test
a system's ability to recover from an error, or in other ways have many
valid execution paths.

The statistical comparisons summarized in Table~\ref{table:comparisons}
suggest that the minimal evidence strategy performs poorly compared to the
median and maximal evidence strategies. This is especially
pronounced for effort reduction, where the Vargha-Delaney
metric scores well over 80\% in favor of the maximal and median
strategy. For recall, the difference between the minimum strategy and
the other variants is small, albeit statistically significant. Furthermore,
the jump from minimal evidence to median evidence is much more
pronounced than the jump from median evidence to maximal evidence.
For effort reduction, there is in fact no statistically discernible
difference between the median and maximal strategies. For recall, the maximal
strategies seems a tiny bit better, but the $A_{12}$ measure suggests the
magnitude of the difference to be small.

Overall, SBLD seems to benefit from extra data, especially additional passing
logs. Failing logs also help, but depend on a proportional amount of passing
logs for SBLD to fully benefit. 
The performance increase from going from minimal data to some data is more pronounced than going from some data to
maximal data. This suggests that there may be diminishing returns to
collecting extra logs, but our investigation cannot prove or disprove this.

\subsection{RQ3: SBLD versus simple pattern-search}

In terms of effort reduction, Table~\ref{table:comparisons} shows that
the pattern-based search clearly beats the minimal evidence variant of
SBLD. It does not, however, beat the median and maximal variants: The
comparison to median evidence suggests a statistically significant win
in favor of median evidence, but the effect reported by $A_{12}$ is
so small that it is unlikely to matter in practice. No statistically
significant difference could be established between the pattern-based
search and SBLD with maximal evidence.

In one sense, it is to be expected that the pattern-based search does
well on effort reduction assuming that events containing words like
"fault" and "error" are rare. The fact that the pattern-based search
works so well could indicate that \CiscoNorway{our industrial partner}
has a well-designed logging infrastructure where such words are
rare and occur at relevant positions in the logs. On the other
hand, it is then notable that the median and maximum variants of SBLD perform
comparably on effort reduction without having any concept of the textual
content in the events.

In terms of recall, however, pattern-based search beats all variants of
SBLD in a statistically significant manner, where the effect size of the
differences is small to medium.  One likely explanation for this better performance is that the
pattern-based search performs very well on Test 43, which SBLD generally
performs less well on. Since the comparisons are run per failing log and test
43 constitutes 29\% of the failing logs (specifically, 267 out of 910 logs), the
performance of test 43 has a massive impact. We return to test 43 and its
impact on our results in Section~\ref{sec:testfourtythree}.

On the whole, SBLD performs similarly to pattern-based search, obtaining
slightly poorer results on recall for reasons that are likely due
to a particular test we discuss below. At any rate, there is no
contradiction in combining SBLD with a traditional pattern-based search.
Analysts could start by issuing a set of pattern-based searches and
run SBLD afterward if the pattern search returned unhelpful results.
Indeed, an excellent and intended use of SBLD is to suggest candidate
signature patterns that, once proven reliable, can be incorporated in a
regular-expression based search to automatically identify known issues
in future runs.

\subsection{What happens in Test 43?}
\label{sec:testfourtythree}

SBLD's performance is much worse on Test 43 than the other tests, which
warrants a dedicated investigation. The first thing we observed in the
results for Test 43 is that all of the ground-truth-relevant events
occurred \emph{exclusively} in failing logs and were often singular
(11 out of the 33) or infrequent (30 out of 33 events occurred in 10\%
of the failing logs or fewer). Consequently, we observed a strong
performance from the \emph{Tarantula} and \emph{Failed only}-measures
that put a high premium on failure-exclusive events. Most of the
interestingness measures, on the other hand, will prefer an event that
is very frequent in the failing logs and sometimes occur in passing logs
over a very rare event that only occurs in failing logs. This goes a
long way in explaining the poor performance on recall. The abundance of
singular events might also suggest that there is an error in the event
abstraction framework, where several events that should be treated as
instances of the same abstract event are treated as separate events. We
discuss this further in Section~\ref{sec:ttv}.

\begin{sloppypar}%
Another observation we made is that the failing logs contained only \emph{two}
ground-truth relevant events, which means that the recorded recall can quickly
fluctuate between $0$, $0.5$ and $1$.
\end{sloppypar}

Would the overall performance improve by retrieving an additional
cluster? A priori, retrieving an extra cluster would strictly improve
or not change recall since more events are retrieved without removing
the previously retrieved events. Furthermore, retrieving an additional
cluster necessarily decreases the effort reduction. We re-ran the
analysis on Test 43 and collected effort reduction and recall scores
for SBLD when retrieving \emph{two} clusters, and found that the added
cluster increased median recall from $0$ to $0.5$ while the median
effort reduction decreased from $0.97$ to $0.72$. While the proportional
increase in recall is larger than the decrease in effort reduction,
this should in our view not be seen as an improvement: As previously
mentioned, the failing logs in this set contain only two ground-truth
relevant events and thus recall is expected to fluctuate greatly.
Secondly, an effort reduction of $0.72$ implies that you still have to
manually inspect 28\% of the data, which in most information retrieval
contexts is unacceptable. An unfortunate aspect of our analysis in this
regard is that we do not account for event \emph{lengths}: An abstracted
event is treated as one atomic entity, but could in reality vary from a
single line to a stack trace that spans several pages. A better measure
of effort reduction should incorporate a notion of event length to
better reflect the real-world effect of retrieving more events.

All in all, Test 43 exhibits a challenge that SBLD is not suited for:
It asks SBLD to prioritize rare events that are exclusive to failing
logs over events that frequently occur in failing logs but might
occasionally occur in passing logs. The majority of interestingness
measures supported by SBLD would prioritize the latter category of
events. In a way, this might suggest that SBLD is not suited for finding
\emph{outliers} and rare events: Rather, it is useful for finding
events that are \emph{characteristic} for failures that have occurred
several times - a "recurring suspect", if you will. An avenue for future
research is to explore ways of letting the user combine a search for
"recurring suspects" with the search for outliers.

\section{Related Work}
\label{sec:relwork}

We distinguish two main lines of related work: 
First, there is other work aimed at automated analysis of log files, 
i.e., our problem domain,
and second, there is other work that shares similarities with our technical approach, 
i.e., our solution domain.

\head{Automated log analysis}
Automated log analysis originates in \emph{system and network monitoring} for security and administration~\cite{lin1990:error,Oliner2007}, 
and saw a revival in recent years due to the needs of \emph{modern software development}, \emph{CE} and \emph{DevOps}~\cite{Hilton2017,Laukkanen2017,Debbiche2014,Olsson2012,Shahin2017,candido2019:contemporary}.

A considerable amount of research has focused on automated \emph{log parsing} or \emph{log abstraction}, 
which aims to reduce and organize log data by recognizing latent structures or templates in the events in a log~\cite{zhu2019:tools,el-masri2020:systematic}.
He et al. analyze the quality of these log parsers and conclude that many of them are not accurate or efficient enough for parsing the logs of modern software systems~\cite{he2018:automated}.
In contrast to these automated approaches, 
our study uses a handcrafted log abstracter developed by \CiscoNorway{our industrial collaborator}.

\emph{Anomaly detection} has traditionally been used for intrusion detection and computer security~\cite{liao2013:intrusion,ramaki2016:survey,ramaki2018:systematic}.
Application-level anomaly detection has been investigated for troubleshooting~\cite{chen2004:failure,zhang2019:robust},
and to assess compliance with service-level agreements~\cite{banerjee2010:logbased,He2018,sauvanaud2018:anomaly}.
Gunter et al. present an infrastructure for troubleshooting of large distributed systems, %
by first (distributively) summarizing high volume event streams before submitting those summaries to a centralized anomaly detector. 
This helps them achieve the fidelity needed for detailed troubleshooting, 
without suffering from the overhead that such detailed instrumentation would bring~\cite{Gunter2007}.
Deeplog by Du et al. enables execution-path and performance anomaly detection in system logs by training a Long Short-Term Memory neural network of the system's expected behavior from the logs, and using that model to flag events and parameter values in the logs that deviate from the model's expectations~\cite{Du2017}.
Similarly, LogRobust by Zhang et al. performs anomaly detection using a bi-LSTM neural network but also detects events that are likely evolved versions of previously seen events, making the learned model more robust to updates in the target logging infrastructure~\cite{zhang2019:robust}.

In earlier work, we use \emph{log clustering} to reduce the effort needed to process a backlog of failing CE logs 
by grouping those logs that failed for similar reasons~\cite{rosenberg2018:use,rosenberg:2018:improving}. 
They build on earlier research that uses log clustering to identify problems in system logs~\cite{Lin2016,Shang2013}.
Common to these approaches is how the contrast between passing and failing logs is used to improve accuracy, 
which is closely related to how SBLD highlights failure-relevant events.

Nagarash et al.~\cite{nagaraj:2012} explore the use of dependency networks to exploit the contrast between two sets of logs, 
one with good and one with bad performance, 
to help developers understand which component(s) likely contain the root cause of performance issues.

An often-occurring challenge is the need to (re)construct an interpretable model of a system's execution.
To this end, several authors investigate the combination of log analysis with (static) source code analysis, 
where they try to (partially) match events in logs to log statements in the code, 
and then use these statements to reconstruct a path through the source code to help determine 
what happened in a failed execution~\cite{Xu2009,yuan:2010:sherlog,zhao2014:lprof,schipper2019:tracing}.
Gadler et al. employ Hidden Markov Models to create a model of a system's usage patterns from logged events~\cite{gadler2017:mining}, while
Pettinato et al. model and analyze the behavior of a complex telescope system using Latent Dirichlet Allocation~\cite{pettinato2019:log}.

Other researchers have analyzed the logs for successful and failing builds, 
to warn for anti-patterns and decay~\cite{vassallo2019:automated}, 
give build repair hints~\cite{Vassallo2018}, 
and automatically repair build scripts~\cite{hassan2018:hirebuild, tarlow2019:learning}. 
Opposite to our work,
these techniques exploit the \emph{overlap} in build systems used by many projects to mine patterns that hint at decay or help repair a failing build, 
whereas we exploit the \emph{contrast} with passing runs for the same project to highlight failure-relevant events.

\begin{sloppypar}
\head{Fault Localization} 
As mentioned, our approach was inspired by Spectrum-Based Fault Localization (SBFL), 
where the fault-proneness of a statement is computed as a function of 
the number of times that the statement was executed in a failing test case, combined with 
the number of times that the statement was skipped in a passing test case~\cite{Jones2002,Chen2002,Abreu2007,Abreu2009,Naish2011}.
This more or less directly translates to the inclusion or exclusion of events in failing, resp. passing logs, 
where the difference is that SBLD adds clustering of the results to enable step-wise presentation of results to the user. 
\end{sloppypar}

A recent survey of Software Fault Localization includes the SBFL literature up to 2014~\cite{Wong2016}.
De Souza et. all extend this with SBFL work up to to 2017, and add an overview of seminal work on automated debugging from 1950 to 1977~\cite{deSouza2017}.
By reflecting on the information-theoretic foundations of fault localization, Perez proposes the DDU metric, 
which can be used to evaluate test suites and predict their diagnostic performance when used in SBFL~\cite{Perez2018}. 
One avenue for future work is exploring how a metric like this can be adapted to our context, 
and see if helps to explain what happened with test 43.

A recent evaluation of \emph{pure} SBFL on large-scale software systems found that it under-performs in these situations 
(only 33-40\% of the bugs are identified with the top 10 of ranked results~\cite{heiden2019:evaluation}. 
The authors discuss several directions beyond pure SBFL, such as combining it with dynamic program analysis techniques, 
including additional text analysis/IR techniques~\cite{Wang2015a}, mutation based fault localization, 
and using SBFL in an interactive feedback-based process, such as whyline-debugging~\cite{ko2008:debugging}.
Pure SBFL is closely related to the Spectrum-Based Log Diagnosis proposed here, 
so we may see similar challenges (in fact, test 43 may already show some of this). 
Of the proposed directions to go beyond pure SBFL, 
both the inclusion of additional text analysis/IR techniques, 
and the application of Spectrum-Based Log Diagnosis in an interactive feedback-based process
are plausible avenues to extend our approach. 
Closely related to the latter option, 
de Souza et al.~\cite{deSouza2018b} assess guidance and filtering strategies to \emph{contextualize} the fault localization process.
Their results suggest that contextualization by guidance and filtering can improve the effectiveness of SBFL,
by classifying more actual bugs in the top ranked results.

 %
\section{Threats to Validity}
\label{sec:ttv}

\head{Construct Validity} %
The signatures that provide our ground truth were devised to determine whether a given log \emph{in its entirety} showed symptoms of a known error.
As discussed in Section~\ref{sec:dataset}, we have used these signatures to detect events that give sufficient evidence for a symptom, 
but there may be other events that could be useful to the user that are not part of our ground truth.
We also assume that the logs exhibit exactly the failures described by the signature expression.
In reality, the logs could contain symptoms of multiple failures beyond the ones described by the signature.

Furthermore, we currently do not distinguish between events that consist of single line of text, 
or events that contain a multi-line stack-trace, although these clearly represent different comprehension efforts.
This threat could be addressed by tracking the \emph{length} of the event contents, 
and using it to further improve the accuracy of our effort reduction measure.

The choice of clustering algorithm and parameters affects the events retrieved, 
but our investigation currently only considers HAC with complete linkage.
While we chose complete linkage to favor compact clusters, 
outliers in the dataset could cause unfavorable clustering outcomes.
Furthermore, using the uncorrected sample standard deviation as threshold criterion 
may be too lenient if the variance in the scores is high.
This threat could be addressed by investigate alternative cluster algorithm and parameter choices.

Moreover, as for the majority of log analysis frameworks, the performance of SBLD strongly depends on the quality of log abstraction. 
An error in the abstraction will directly propagate to SBLD: 
For example, if abstraction fails to identify two concrete events as being instances of the same generic event, 
their aggregated frequencies will be smaller and consequently treated as less interesting by SBLD.
Similarly, the accuracy will suffer if two events that represent distinct generic events are treated as instances of the same generic event.
Future work could investigate alternative log abstraction approaches.

\head{Internal Validity} %
While our heatmaps illustrate the interaction between additional data and SBLD performance, 
they are not sufficient to prove a causal relationship between performance and added data.
Our statistical comparisons suggests that a strategy of maximizing data is generally preferable, 
but they are not sufficient for discussing the respective contribution of failing or passing logs.

\head{External Validity} %
This investigation is concerned with a single dataset from one industrial partner.
Studies using additional datasets from other contexts is needed to assess the generalizability of SBLD to other domains.
Moreover, while SBLD is made to help users diagnose problems that are not already well understood,
we are assessing it on a dataset of \emph{known} problems.
It could be that these errors, being known, are of a kind that are generally easier to identify than most errors.
Studying SBLD in-situ over time and directly assessing whether end users found it helpful
in diagnosis would better indicate the generalizability of our approach.

\section{Concluding Remarks}
\label{sec:conclusion}

\head{Contributions}
This paper presents and evaluates Spectrum-Based Log Diagnosis (SBLD), 
a method for automatically identifying segments of failing logs 
that are likely to help users diagnose failures. 
Our empirical investigation of SBLD addresses the following questions: 
(i) How well does SBLD reduce the \emph{effort needed} to identify all \emph{failure-relevant events} in the log for a failing run? 
(ii) How is the \emph{performance} of SBLD affected by \emph{available data}? 
(iii) How does SBLD compare to searching for \emph{simple textual patterns} that often occur in failure-relevant events? 

\head{Results}
In response to (i), 
we find that SBLD generally retrieves the failure-relevant events in a compact manner 
that effectively reduces the effort needed to identify failure-relevant events. 
In response to (ii), 
we find that SBLD benefits from addition data, especially more logs from successful runs. 
SBLD also benefits from additional logs from failing runs if there is a proportional amount of successful runs in the set. 
We also find that the effect of added data is most pronounced when going from little data to \emph{some} data rather than from \emph{some} data to maximal data. 
In response to (iii), 
we find that SBLD achieves roughly the same effort reduction as traditional search-based methods but obtains slightly lower recall. 
We trace the likely cause of this discrepancy on recall to a prominent part of our dataset, whose ground truth emphasizes rare events. 
A lesson learned in this regard is that SBLD is not suited for finding statistical outliers but rather \emph{recurring suspects} 
that characterize the observed failures. 
Furthermore, the investigation highlights that traditional pattern-based search and SBLD can complement each other nicely: 
Users can resort to SBLD if they are unhappy with what the pattern-based searches turn
up, and SBLD is an excellent method for finding characteristic textual patterns
that can form the basis of automated failure identification methods.

\head{Conclusions}
We conclude that SBLD shows promise as a method diagnosing failing runs, 
that its performance is positively affected by additional data, 
but that it does not outperform textual search on the dataset considered. 

\head{Future work}
We see the following directions for future work: 
(a) investigate SBLD's performance on other datasets, to better assess generalizability, 
(b) explore the impact of alternative log abstraction mechanisms,
(c) explore ways of combining SBLD with outlier detection, to accommodate different user needs, 
(d) adapt the Perez' DDU metric to our context and see if it can help predict diagnostic efficiency,
(e) experiment with extensions of \emph{pure SBLD} that include additional text analysis/IR techniques, 
    or apply it in an interactive feedback-based process
(f) rigorously assess (extensions of) SBLD in in-situ experiments.

\begin{acks}
We thank Marius Liaaen and Thomas Nornes of Cisco Systems Norway for help with obtaining and understanding the dataset, for developing the log abstraction
mechanisms and for extensive discussions.
This work is supported by the \grantsponsor{RCN}{Research Council of Norway}{https://www.rcn.no} through the
Certus SFI (\grantnum{RCN}{\#203461/030)}.
The empirical evaluation was performed on resources provided by \textsc{uninett s}igma2,
the national infrastructure for high performance computing and data
storage in Norway.
\end{acks}

 \printbibliography

@ARTICLE{Laukkanen2017,
  AUTHOR = {Laukkanen, Eero and Itkonen, Juha and Lassenius, Casper},
  DATE = {2017-02},
  DOI = {10.1016/j.infsof.2016.10.001},
  ISSN = {09505849},
  JOURNALTITLE = {Information and Software Technology},
  PAGES = {55--79},
  TITLE = {Problems, Causes and Solutions When Adopting Continuous Delivery - {{A}} Systematic Literature Review},
  VOLUME = {82},
}

@INPROCEEDINGS{Hilton2017,
  AUTHOR = {Hilton, Michael and Nelson, Nicholas and Tunnell, Timothy and Marinov, Darko and Dig, Danny},
  LOCATION = {New York, New York, USA},
  PUBLISHER = {ACM},
  BOOKTITLE = {European {{Software Engineering Conference}} and {{ACM SIGSOFT Symposium}} on the {{Foundations}} of {{Software Engineering}} ({{ESEC}}/{{FSE}})},
  DATE = {2017},
  DOI = {10.1145/3106237.3106270},
  ISBN = {978-1-4503-5105-8},
  PAGES = {197--207},
  TITLE = {Trade-Offs in Continuous Integration: Assurance, Security, and Flexibility},
}

@ARTICLE{Shahin2017,
  AUTHOR = {Shahin, Mojtaba and Ali Babar, Muhammad and Zhu, Liming},
  DATE = {2017},
  DOI = {10.1109/ACCESS.2017.2685629},
  ISSN = {2169-3536},
  JOURNALTITLE = {IEEE Access},
  PAGES = {3909--3943},
  TITLE = {Continuous {{Integration}}, {{Delivery}} and {{Deployment}}: {{A Systematic Review}} on {{Approaches}}, {{Tools}}, {{Challenges}} and {{Practices}}},
  VOLUME = {5},
}

@INPROCEEDINGS{Debbiche2014,
  AUTHOR = {Debbiche, Adam and Dien{é}r, Mikael and Berntsson Svensson, Richard},
  PUBLISHER = {Springer},
  BOOKTITLE = {International {{Conference}} on {{Product}}-{{Focused Software Process Improvement}} ({{PROFES}})},
  DATE = {2014},
  DOI = {10.1007/978-3-319-13835-0_2},
  PAGES = {17--32},
  SERIES = {{{LNCS}}},
  TITLE = {Challenges {{When Adopting Continuous Integration}}: {{A Case Study}}},
  VOLUME = {8892},
}

@INPROCEEDINGS{Olsson2012,
  AUTHOR = {Olsson, Helena Holmstrom and Alahyari, Hiva and Bosch, Jan},
  PUBLISHER = {IEEE},
  BOOKTITLE = {Euromicro {{Conference}} on {{Software Engineering}} and {{Advanced Applications}}},
  DATE = {2012-09},
  DOI = {10.1109/SEAA.2012.54},
  ISBN = {978-0-7695-4790-9},
  PAGES = {392--399},
  TITLE = {Climbing the {{Stairway}} to {{Heaven}} - {{A Mulitiple}}-{{Case Study Exploring Barriers}} in the {{Transition}} from {{Agile Development}} towards {{Continuous Deployment}} of {{Software}}},
}

@INPROCEEDINGS{Jones2002,
  AUTHOR = {Jones, James A. and Harrold, Mary Jean and Stasko, John},
  BOOKTITLE = {{{IEEE}}/{{ACM International Conference}} on {{Software Engineering}} ({{ICSE}})},
  DATE = {2002},
  DOI = {10.1145/581339.581397},
  PAGES = {467},
  TITLE = {Visualization of Test Information to Assist Fault Localization},
}

@INPROCEEDINGS{Abreu2007,
  AUTHOR = {Abreu, Rui and Zoeteweij, Peter and {van Gemund}, Arjan J.C.},
  PUBLISHER = {IEEE},
  BOOKTITLE = {Testing: {{Academic}} and {{Industrial Conference Practice}} and {{Research Techniques}} ({{TAICPART}})},
  DATE = {2007-09},
  DOI = {10.1109/TAIC.PART.2007.13},
  ISBN = {0-7695-2984-4},
  PAGES = {89--98},
  TITLE = {On the {{Accuracy}} of {{Spectrum}}-Based {{Fault Localization}}},
}

@ARTICLE{Abreu2009,
  AUTHOR = {Abreu, Rui and Zoeteweij, Peter and Golsteijn, Rob and {van Gemund}, Arjan J C},
  PUBLISHER = {Elsevier},
  DATE = {2009},
  DOI = {10.1016/j.jss.2009.06.035},
  ISSN = {0164-1212},
  JOURNALTITLE = {Journal of Systems and Software},
  NUMBER = {11},
  PAGES = {1780--1792},
  TITLE = {A Practical Evaluation of Spectrum-Based Fault Localization},
  VOLUME = {82},
}

@ARTICLE{Wong2016,
  AUTHOR = {Wong, W. Eric and Gao, Ruizhi and Li, Yihao and Abreu, Rui and Wotawa, Franz},
  DATE = {2016},
  DOI = {10.1109/TSE.2016.2521368},
  ISSN = {0098-5589},
  JOURNALTITLE = {IEEE Transactions on Software Engineering},
  NUMBER = {8},
  TITLE = {A {{Survey}} on {{Software Fault Localization}}},
  VOLUME = {42},
}

@REPORT{Yoo2014,
  AUTHOR = {Yoo, Shin. and Xie, Xiaoyuan and Kuo, Fei-Ching and Chen, Tsong Yueh and Harman, Mark},
  INSTITUTION = {UCL Department of Computer Science},
  DATE = {2014},
  NUMBER = {RN/14/14},
  TITLE = {No {{Pot}} of {{Gold}} at the {{End}} of {{Program Spectrum Rainbow}}: {{Greatest Risk Evaluation Formula Does Not Exist}}},
  TYPE = {techreport},
}

@INPROCEEDINGS{renieres2003:fault,
  AUTHOR = {Renieres, M. and Reiss, S. P.},
  BOOKTITLE = {{{IEEE}}/{{ACM International Conference}} on {{Automated Software Engineering}} ({{ASE}})},
  DATE = {2003-10},
  DOI = {10.1109/ASE.2003.1240292},
  ISSN = {1938-4300},
  PAGES = {30--39},
  TITLE = {Fault Localization with Nearest Neighbor Queries},
}

@INPROCEEDINGS{Jones2001,
  AUTHOR = {Jones, James A. and Harrold, Mary Jean and Stasko, John},
  BOOKTITLE = {{{ICSE Workshop}} on {{Software Visualization}}},
  DATE = {2001},
  PAGES = {5},
  TITLE = {Visualization for Fault Localization},
}

@ARTICLE{Jaccard1912,
  AUTHOR = {Jaccard, Paul},
  DATE = {1912-02},
  DOI = {10.1111/j.1469-8137.1912.tb05611.x},
  ISSN = {0028-646X},
  JOURNALTITLE = {New Phytologist},
  NUMBER = {2},
  PAGES = {37--50},
  TITLE = {The {{Distribution}} of the {{Flora}} in the {{Alpine Zone}}},
  VOLUME = {11},
}

@INPROCEEDINGS{Chen2002,
  AUTHOR = {Chen, M.Y. and Kiciman, Emre and Fratkin, Eugene and Fox, Armando and Brewer, Eric},
  PUBLISHER = {IEEE},
  BOOKTITLE = {{{IEEE}}/{{IFIP International Conference}} on {{Dependable Systems}} and {{Networks}} ({{DSN}})},
  DATE = {2002},
  DOI = {10.1109/DSN.2002.1029005},
  ISBN = {0-7695-1597-5},
  PAGES = {595--604},
  TITLE = {Pinpoint: Problem Determination in Large, Dynamic {{Internet}} Services},
}

@ARTICLE{Ochiai1957,
  AUTHOR = {Ochiai, Akira},
  DATE = {1957},
  DOI = {10.2331/suisan.22.526},
  ISSN = {1349-998X},
  JOURNALTITLE = {Nippon Suisan Gakkaishi},
  NUMBER = {9},
  PAGES = {526--530},
  TITLE = {Zoogeographical {{Studies}} on the {{Soleoid Fishes Found}} in {{Japan}} and Its {{Neighhouring Regions}}-{{II}}},
  VOLUME = {22},
}

@INPROCEEDINGS{Abreu2006,
  AUTHOR = {Abreu, Rui and Zoeteweij, Peter and Van Gemund, Arjan},
  PUBLISHER = {IEEE},
  BOOKTITLE = {Pacific {{Rim International Symposium}} on {{Dependable Computing}} ({{PRDC}})},
  DATE = {2006},
  DOI = {10.1109/PRDC.2006.18},
  ISBN = {0-7695-2724-8},
  PAGES = {39--46},
  TITLE = {An {{Evaluation}} of {{Similarity Coefficients}} for {{Software Fault Localization}}},
}

@ARTICLE{Naish2011,
  AUTHOR = {Naish, Lee and Lee, Hua Jie and Ramamohanarao, Kotagiri},
  DATE = {2011-08},
  DOI = {10.1145/2000791.2000795},
  JOURNALTITLE = {ACM Transactions on Software Engineering and Methodology},
  NUMBER = {3},
  PAGES = {1--32},
  TITLE = {A Model for Spectra-Based Software Diagnosis},
  VOLUME = {20},
}

@THESIS{Gonzalez2007,
  AUTHOR = {{Gonz{á}lez-S{á}nchez}, Alberto},
  INSTITUTION = {Delft University of Technology, The Netherlands},
  DATE = {2007},
  TITLE = {Automatic Error Detection Techniques Based on Dynamic Invariants},
  TYPE = {{{MSc}}. Thesis},
}

@ARTICLE{Wong2014,
  AUTHOR = {Wong, W. Eric and Debroy, Vidroha and Gao, Ruizhi and Li, Yihao},
  DATE = {2014-03},
  DOI = {10.1109/TR.2013.2285319},
  ISSN = {0018-9529},
  JOURNALTITLE = {IEEE Transactions on Reliability},
  NUMBER = {1},
  PAGES = {290--308},
  TITLE = {The {{DStar Method}} for {{Effective Software Fault Localization}}},
  VOLUME = {63},
}

@INPROCEEDINGS{Wong2007,
  AUTHOR = {Wong, W. Eric and Qi, Yu and Zhao, Lei and Cai, Kai-Yuan},
  PUBLISHER = {IEEE},
  BOOKTITLE = {Annual {{International Computer Software}} and {{Applications Conference}} ({{COMPSAC}})},
  DATE = {2007-07},
  DOI = {10.1109/COMPSAC.2007.109},
  ISBN = {0-7695-2870-8},
  PAGES = {449--456},
  TITLE = {Effective {{Fault Localization}} Using {{Code Coverage}}},
  VOLUME = {1},
}

@ARTICLE{Wong2010,
  AUTHOR = {Wong, W. Eric and Debroy, Vidroha and Choi, Byoungju},
  DATE = {2010-02},
  DOI = {10.1016/j.jss.2009.09.037},
  ISSN = {0164-1212},
  JOURNALTITLE = {Journal of Systems and Software},
  NUMBER = {2},
  PAGES = {188--208},
  TITLE = {A Family of Code Coverage-Based Heuristics for Effective Fault Localization},
  VOLUME = {83},
}

@ARTICLE{Kulczynski1927,
  AUTHOR = {Kulczy{ń}ski, Stanis{ł}aw},
  DATE = {1927},
  JOURNALTITLE = {Bulletin International de l'Academie Polonaise des Sciences et des Lettres, Classe des Sciences Mathematiques et Naturelles, B (Sciences Naturelles)},
  PAGES = {57--203},
  TITLE = {Die {{Pflanzenassoziationen}} Der {{Pieninen}}},
  VOLUME = {II},
}

@BOOK{manning2008introduction,
  AUTHOR = {Manning, Christopher D and Raghavan, Prabhakar and Sch{ü}tze, Hinrich and others},
  PUBLISHER = {Cambridge University Press},
  DATE = {2008},
  ISBN = {978-0-521-86571-5},
  TITLE = {Introduction to {{Information Retrieval}}},
}

@ARTICLE{2020SciPy-NMeth,
  AUTHOR = {Virtanen, Pauli and Gommers, Ralf and Oliphant, Travis E. and Haberland, Matt and Reddy, Tyler and Cournapeau, David and Burovski, Evgeni and Peterson, Pearu and Weckesser, Warren and Bright, Jonathan and {van der Walt}, St{é}fan J. and Brett, Matthew and Wilson, Joshua and Jarrod Millman, K. and Mayorov, Nikolay and Nelson, Andrew R. J. and Jones, Eric and Kern, Robert and Larson, Eric and Carey, CJ and Polat, {İ}lhan and Feng, Yu and Moore, Eric W. and {Vand erPlas}, Jake and Laxalde, Denis and Perktold, Josef and Cimrman, Robert and Henriksen, Ian and Quintero, E. A. and Harris, Charles R and Archibald, Anne M. and Ribeiro, Ant{ô}nio H. and Pedregosa, Fabian and {van Mulbregt}, Paul and Contributors, SciPy 1. 0},
  DATE = {2020},
  DOI = {10.1038/s41592-019-0686-2},
  JOURNALTITLE = {Nature Methods},
  PAGES = {261--272},
  TITLE = {{{SciPy}} 1.0: {{Fundamental}} Algorithms for Scientific Computing in Python},
  VOLUME = {17},
}

@ARTICLE{wilcoxon1945,
  AUTHOR = {Wilcoxon, Frank},
  DATE = {1945},
  DOI = {10.2307/3001968},
  ISSN = {00994987},
  JOURNALTITLE = {Biometrics Bulletin},
  NUMBER = {6},
  PAGES = {80--83},
  TITLE = {Individual {{Comparisons}} by {{Ranking Methods}}},
  VOLUME = {1},
}

@ARTICLE{Pratt1959,
  AUTHOR = {Pratt, John W.},
  DATE = {1959-09},
  DOI = {10.2307/2282543},
  ISSN = {01621459},
  JOURNALTITLE = {Journal of the American Statistical Association},
  NUMBER = {287},
  PAGES = {655},
  TITLE = {Remarks on {{Zeros}} and {{Ties}} in the {{Wilcoxon Signed Rank Procedures}}},
  VOLUME = {54},
}

@ARTICLE{Holm1979,
  AUTHOR = {Holm, Sture},
  DATE = {1979},
  JOURNALTITLE = {Scandinavian Journal of Statistics},
  NUMBER = {2},
  PAGES = {65--70},
  TITLE = {A {{Simple Sequentially Rejective Multiple Test Procedure}}},
  VOLUME = {6},
}

@ARTICLE{Vargha2000,
  AUTHOR = {Vargha, Andr{á}s and Delaney, Harold D.},
  DATE = {2000-06},
  DOI = {10.3102/10769986025002101},
  ISSN = {1076-9986},
  JOURNALTITLE = {Journal of Educational and Behavioral Statistics},
  NUMBER = {2},
  PAGES = {101--132},
  TITLE = {A {{Critique}} and {{Improvement}} of the {{CL Common Language Effect Size Statistics}} of {{McGraw}} and {{Wong}}},
  VOLUME = {25},
}

@ARTICLE{lin1990:error,
  AUTHOR = {Lin, T.-T.Y. and Siewiorek, D.P.},
  DATE = {1990-10},
  DOI = {10.1109/24.58720},
  ISSN = {1558-1721},
  JOURNALTITLE = {IEEE Transactions on Reliability},
  NUMBER = {4},
  PAGES = {419--432},
  SHORTTITLE = {Error Log Analysis},
  TITLE = {Error Log Analysis: Statistical Modeling and Heuristic Trend Analysis},
  VOLUME = {39},
}

@INPROCEEDINGS{Oliner2007,
  AUTHOR = {Oliner, Adam and Stearley, Jon},
  BOOKTITLE = {{{IEEE}}/{{IFIP International Conference}} on {{Dependable Systems}} and {{Networks}} ({{DSN}})},
  DATE = {2007},
  DOI = {10.1109/DSN.2007.103},
  ISBN = {0-7695-2855-4},
  PAGES = {575--584},
  TITLE = {What Supercomputers Say: {{A}} Study of Five System Logs},
}

@REPORT{candido2019:contemporary,
  AUTHOR = {Candido, Jeanderson and Aniche, Maur{í}cio and {van Deursen}, Arie},
  INSTITUTION = {CoRR e-print},
  LANGUAGE = {en},
  DATE = {2019-12},
  EPRINT = {1912.05878},
  EPRINTTYPE = {arxiv},
  TITLE = {Contemporary {{Software Monitoring}}: {{A Systematic Literature Review}}},
  TYPE = {{{arXiv}}: 1912.05878v1},
}

@INPROCEEDINGS{zhu2019:tools,
  AUTHOR = {Zhu, Jieming and He, Shilin and Liu, Jinyang and He, Pinjia and Xie, Qi and Zheng, Zibin and Lyu, Michael R.},
  LANGUAGE = {en},
  LOCATION = {Montreal, QC, Canada},
  PUBLISHER = {IEEE},
  BOOKTITLE = {{{IEEE}}/{{ACM International Conference}} on {{Software Engineering}}: {{Software Engineering}} in {{Practice}} ({{ICSE}}-{{SEIP}})},
  DATE = {2019-05},
  DOI = {10.1109/ICSE-SEIP.2019.00021},
  ISBN = {978-1-72811-760-7},
  PAGES = {121--130},
  TITLE = {Tools and {{Benchmarks}} for {{Automated Log Parsing}}},
}

@ARTICLE{el-masri2020:systematic,
  AUTHOR = {{El-Masri}, Diana and Petrillo, Fabio and Gu{é}h{é}neuc, Yann-Ga{ë}l and {Hamou-Lhadj}, Abdelwahab and Bouziane, Anas},
  LANGUAGE = {en},
  DATE = {2020-06},
  DOI = {10.1016/j.infsof.2020.106276},
  ISSN = {0950-5849},
  JOURNALTITLE = {Information and Software Technology},
  PAGES = {106276},
  TITLE = {A Systematic Literature Review on Automated Log Abstraction Techniques},
  VOLUME = {122},
}

@ARTICLE{he2018:automated,
  AUTHOR = {He, Pinjia and Zhu, Jieming and He, Shilin and Li, Jian and Lyu, Michael R.},
  DATE = {2018-11},
  DOI = {10.1109/TDSC.2017.2762673},
  ISSN = {1941-0018},
  JOURNALTITLE = {IEEE Transactions on Dependable and Secure Computing},
  NUMBER = {6},
  PAGES = {931--944},
  TITLE = {Towards {{Automated Log Parsing}} for {{Large}}-{{Scale Log Data Analysis}}},
  VOLUME = {15},
}

@ARTICLE{liao2013:intrusion,
  AUTHOR = {Liao, H.-J. and Richard Lin, C.-H. and Lin, Y.-C. and Tung, K.-Y.},
  DATE = {2013},
  DOI = {10.1016/j.jnca.2012.09.004},
  JOURNALTITLE = {Journal of Network and Computer Applications},
  NUMBER = {1},
  PAGES = {16--24},
  SHORTTITLE = {Intrusion Detection System},
  TITLE = {Intrusion Detection System: {{A}} Comprehensive Review},
  VOLUME = {36},
}

@ARTICLE{ramaki2016:survey,
  AUTHOR = {Ramaki, A.A. and Atani, R.E.},
  DATE = {2016},
  DOI = {10.1002/sec.1647},
  JOURNALTITLE = {Security and Communication Networks},
  NUMBER = {17},
  PAGES = {4751--4776},
  SHORTTITLE = {A Survey of {{IT}} Early Warning Systems},
  TITLE = {A Survey of {{IT}} Early Warning Systems: Architectures, Challenges, and Solutions},
  VOLUME = {9},
}

@ARTICLE{ramaki2018:systematic,
  AUTHOR = {Ramaki, A.A. and Rasoolzadegan, A. and Bafghi, A.G.},
  DATE = {2018},
  DOI = {10.1145/3184898},
  JOURNALTITLE = {ACM Computing Surveys},
  NUMBER = {3},
  TITLE = {A Systematic Mapping Study on Intrusion Alert Analysis in Intrusion Detection Systems},
  VOLUME = {51},
}

@INPROCEEDINGS{chen2004:failure,
  AUTHOR = {Chen, M. and Zheng, A.X. and Lloyd, J. and Jordan, M.I. and Brewer, E.},
  BOOKTITLE = {International {{Conference}} on {{Autonomic Computing}}},
  DATE = {2004-05},
  DOI = {10.1109/ICAC.2004.1301345},
  PAGES = {36--43},
  TITLE = {Failure Diagnosis Using Decision Trees},
}

@INPROCEEDINGS{zhang2019:robust,
  AUTHOR = {Zhang, Xu and Xu, Yong and Lin, Qingwei and Qiao, Bo and Zhang, Hongyu and Dang, Yingnong and Xie, Chunyu and Yang, Xinsheng and Cheng, Qian and Li, Ze and Chen, Junjie and He, Xiaoting and Yao, Randolph and Lou, Jian-Guang and Chintalapati, Murali and Shen, Furao and Zhang, Dongmei},
  LOCATION = {New York, NY, USA},
  PUBLISHER = {Association for Computing Machinery},
  BOOKTITLE = {European {{Software Engineering Conference}} and {{ACM SIGSOFT Symposium}} on the {{Foundations}} of {{Software Engineering}} ({{ESEC}}/{{FSE}})},
  DATE = {2019},
  DOI = {10.1145/3338906.3338931},
  ISBN = {978-1-4503-5572-8},
  PAGES = {807--817},
  SERIES = {{{ESEC}}/{{FSE}} 2019},
  TITLE = {Robust Log-Based Anomaly Detection on Unstable Log Data},
}

@INPROCEEDINGS{banerjee2010:logbased,
  AUTHOR = {Banerjee, Sean and Srikanth, Hema and Cukic, Bojan},
  BOOKTITLE = {{{IEEE International Symposium}} on {{Software Reliability Engineering}} ({{ISSRE}})},
  DATE = {2010-11},
  DOI = {10.1109/ISSRE.2010.46},
  ISSN = {2332-6549},
  PAGES = {239--248},
  TITLE = {Log-{{Based Reliability Analysis}} of {{Software}} as a {{Service}} ({{SaaS}})},
}

@INPROCEEDINGS{He2018,
  AUTHOR = {He, Shilin and Lin, Qingwei and Lou, Jian-guang and Zhang, Hongyu and Lyu, Michael R and Zhang, Dongmei},
  LOCATION = {New York, New York, USA},
  PUBLISHER = {ACM Press},
  BOOKTITLE = {European {{Software Engineering Conference}} and {{ACM SIGSOFT Symposium}} on the {{Foundations}} of {{Software Engineering}} ({{ESEC}}/{{FSE}})},
  DATE = {2018},
  DOI = {10.1145/3236024.3236083},
  ISBN = {978-1-4503-5573-5},
  PAGES = {60--70},
  TITLE = {Identifying Impactful Service System Problems via Log Analysis},
}

@ARTICLE{sauvanaud2018:anomaly,
  AUTHOR = {Sauvanaud, C. and Ka{â}niche, M. and Kanoun, K. and Lazri, K. and Da Silva Silvestre, G.},
  DATE = {2018},
  DOI = {10.1016/j.jss.2018.01.039},
  JOURNALTITLE = {Journal of Systems and Software},
  PAGES = {84--106},
  SHORTTITLE = {Anomaly Detection and Diagnosis for Cloud Services},
  TITLE = {Anomaly Detection and Diagnosis for Cloud Services: {{Practical}} Experiments and Lessons Learned},
  VOLUME = {139},
}

@INPROCEEDINGS{Gunter2007,
  AUTHOR = {Gunter, Dan and Tierney, Brian L. and Brown, Aaron and Swany, Martin and Bresnahan, John and Schopf, Jennifer M.},
  BOOKTITLE = {{{IEEE}}/{{ACM International Workshop}} on {{Grid Computing}}},
  DATE = {2007},
  DOI = {10.1109/GRID.2007.4354137},
  ISBN = {1-4244-1560-8},
  PAGES = {226--234},
  TITLE = {Log Summarization and Anomaly Detection for Troubleshooting Distributed Systems},
}

@INPROCEEDINGS{Du2017,
  AUTHOR = {Du, Min and Li, Feifei and Zheng, Guineng and Srikumar, Vivek},
  LOCATION = {New York, New York, USA},
  PUBLISHER = {ACM Press},
  BOOKTITLE = {{{SIGSAC Conference}} on {{Computer}} and {{Communications Security}} ({{CCS}})},
  DATE = {2017},
  DOI = {10.1145/3133956.3134015},
  ISBN = {978-1-4503-4946-8},
  ISSN = {15437221},
  PAGES = {1285--1298},
  TITLE = {{{DeepLog}}: {{Anomaly}} Detection and Diagnosis from System Logs through Deep Learning.},
}

@INPROCEEDINGS{rosenberg2018:use,
  AUTHOR = {Rosenberg, Carl Martin and Moonen, Leon},
  BOOKTITLE = {Asia-{{Pacific Software Engineering Conference}} ({{APSEC}})},
  DATE = {2018-12},
  DOI = {10.1109/APSEC.2018.00032},
  ISSN = {2640-0715},
  PAGES = {179--188},
  TITLE = {On the {{Use}} of {{Automated Log Clustering}} to {{Support Effort Reduction}} in {{Continuous Engineering}}},
}

@INPROCEEDINGS{rosenberg:2018:improving,
  AUTHOR = {Rosenberg, Carl Martin and Moonen, Leon},
  PUBLISHER = {ACM},
  BOOKTITLE = {International {{Symposium}} on {{Empirical Software Engineering}} and {{Measurement}} ({{ESEM}})},
  DATE = {2018},
  DOI = {10.1145/3239235.3239248},
  PAGES = {Article No. 16},
  TITLE = {Improving {{Problem Identification}} via {{Automated Log Clustering}} Using {{Dimensionality Reduction}}},
}

@INPROCEEDINGS{Lin2016,
  AUTHOR = {Lin, Qingwei and Zhang, Hongyu and Lou, Jian-Guang and Zhang, Yu and Chen, Xuewei},
  LOCATION = {New York, NY, USA},
  PUBLISHER = {ACM},
  BOOKTITLE = {{{IEEE}}/{{ACM International Conference}} on {{Software Engineering}}: {{Software Engineering}} in {{Practice}} ({{ICSE}}-{{SEIP}})},
  DATE = {2016},
  DOI = {10.1145/2889160.2889232},
  ISBN = {978-1-4503-4205-6},
  PAGES = {102--111},
  TITLE = {Log {{Clustering Based Problem Identification}} for {{Online Service Systems}}},
}

@INPROCEEDINGS{Shang2013,
  AUTHOR = {Shang, Weiyi and Jiang, Zhen Ming and Hemmati, Hadi and Adams, Brain and Hassan, Ahmed E. and Martin, Patrick},
  PUBLISHER = {IEEE},
  BOOKTITLE = {{{IEEE}}/{{ACM International Conference}} on {{Software Engineering}} ({{ICSE}})},
  DATE = {2013-05},
  DOI = {10.1109/ICSE.2013.6606586},
  ISBN = {978-1-4673-3076-3},
  ISSN = {02705257},
  TITLE = {Assisting Developers of {{Big Data Analytics Applications}} When Deploying on {{Hadoop}} Clouds},
}

@INPROCEEDINGS{nagaraj:2012,
  AUTHOR = {Nagaraj, Karthik and Killian, Charles and Neville, Jennifer},
  BOOKTITLE = {{{USENIX Symposium}} on {{Networked Systems Design}} and {{Implementation}} ({{NSDI}})},
  DATE = {2012},
  PAGES = {353--366},
  TITLE = {Structured {{Comparative Analysis}} of {{Systems Logs}} to {{Diagnose Performance Problems}}},
}

@INPROCEEDINGS{Xu2009,
  AUTHOR = {Xu, Wei and Huang, Ling and Fox, Armando and Patterson, David and Jordan, Michael I.},
  PUBLISHER = {ACM},
  BOOKTITLE = {{{ACM SIGOPS Symposium}} on {{Operating Systems Principles}} ({{SOSP}})},
  DATE = {2009},
  DOI = {10.1145/1629575.1629587},
  ISBN = {978-1-60558-752-3},
  PAGES = {117--132},
  TITLE = {Detecting Large-Scale System Problems by Mining Console Logs},
}

@INPROCEEDINGS{yuan:2010:sherlog,
  AUTHOR = {Yuan, Ding and Mai, Haohui and Xiong, Weiwei and Tan, Lin and Zhou, Yuanyuan and Pasupathy, Shankar},
  BOOKTITLE = {Architectural {{Support}} for {{Programming Languages}} and {{Operating Systems}} ({{ASPLOS}})},
  DATE = {2010},
  DOI = {10.1145/1736020.1736038},
  ISBN = {978-1-60558-839-1},
  PAGES = {143},
  TITLE = {{{SherLog}}: {{Error Diagnosis}} by {{Connecting Clues}} from {{Run}}-Time {{Logs}}},
}

@INPROCEEDINGS{zhao2014:lprof,
  AUTHOR = {Zhao, Xu and Zhang, Yongle and Lion, David and Ullah, Muhammad Faizan and Luo, Yu and Yuan, Ding and Stumm, Michael},
  LANGUAGE = {en},
  BOOKTITLE = {11th \{\vphantom\}{{USENIX}}\vphantom\\ {{Symposium}} on {{Operating Systems Design}} and {{Implementation}} (\{\vphantom\}{{OSDI}}\vphantom\\ 14)},
  DATE = {2014},
  ISBN = {978-1-931971-16-4},
  PAGES = {629--644},
  SHORTTITLE = {Lprof},
  TITLE = {Lprof: {{A Non}}-Intrusive {{Request Flow Profiler}} for {{Distributed Systems}}},
}

@INPROCEEDINGS{schipper2019:tracing,
  AUTHOR = {Schipper, Daan and Aniche, Maur{ı}cio and {van Deursen}, Arie},
  LANGUAGE = {en},
  BOOKTITLE = {{{IEEE}}/{{ACM International Conference}} on {{Mining Software Repositories}} ({{MSR}})},
  DATE = {2019-05},
  DOI = {10.1109/MSR.2019.00081},
  PAGES = {545--549},
  TITLE = {Tracing {{Back Log Data}} to Its {{Log Statement}}: {{From Research}} to {{Practice}}},
}

@INPROCEEDINGS{gadler2017:mining,
  AUTHOR = {Gadler, Daniele and Mairegger, Michael and Janes, Andrea and Russo, Barbara},
  LOCATION = {Markham, Ontario, Canada},
  PUBLISHER = {IEEE Press},
  BOOKTITLE = {Proceedings of the 11th {{ACM}}/{{IEEE International Symposium}} on {{Empirical Software Engineering}} and {{Measurement}}},
  DATE = {2017-11},
  DOI = {10.1109/ESEM.2017.47},
  ISBN = {978-1-5090-4039-1},
  PAGES = {334--343},
  SERIES = {{{ESEM}} '17},
  TITLE = {Mining Logs to Model the Use of a System},
}

@ARTICLE{pettinato2019:log,
  AUTHOR = {Pettinato, Michele and Gil, Juan Pablo and Galeas, Patricio and Russo, Barbara},
  LANGUAGE = {en},
  DATE = {2019-10},
  DOI = {10.1016/j.infsof.2019.06.011},
  ISSN = {0950-5849},
  JOURNALTITLE = {Information and Software Technology},
  PAGES = {121--136},
  SHORTTITLE = {Log Mining to Re-Construct System Behavior},
  TITLE = {Log Mining to Re-Construct System Behavior: {{An}} Exploratory Study on a Large Telescope System},
  VOLUME = {114},
}

@INPROCEEDINGS{vassallo2019:automated,
  AUTHOR = {Vassallo, Carmine},
  LANGUAGE = {en},
  BOOKTITLE = {{{IEEE}}/{{ACM International Conference}} on {{Software Engineering}} ({{ICSE}})},
  DATE = {2019-05},
  PAGES = {105--115},
  TITLE = {Automated {{Reporting}} of {{Anti}}-{{Patterns}} and {{Decay}} in {{Continuous Integration}}},
}

@INPROCEEDINGS{Vassallo2018,
  AUTHOR = {Vassallo, Carmine and Proksch, Sebastian and Zemp, Timothy and Gall, Harald C},
  BOOKTITLE = {{{IEEE International Conference}} on {{Program Comprehension}} ({{ICPC}})},
  DATE = {2018},
  DOI = {10.1145/3196321.3196350},
  ISBN = {978-1-4503-5714-2},
  PAGES = {41--51},
  TITLE = {Un-Break {{My Build}}: {{Assisting Developers}} with {{Build Repair Hints}}},
}

@INPROCEEDINGS{hassan2018:hirebuild,
  AUTHOR = {Hassan, Foyzul and Wang, Xiaoyin},
  LANGUAGE = {en},
  LOCATION = {Gothenburg, Sweden},
  PUBLISHER = {ACM Press},
  BOOKTITLE = {{{IEEE}}/{{ACM International Conference}} on {{Software Engineering}} ({{ICSE}})},
  DATE = {2018},
  DOI = {10.1145/3180155.3180181},
  ISBN = {978-1-4503-5638-1},
  PAGES = {1078--1089},
  SHORTTITLE = {{{HireBuild}}},
  TITLE = {{{HireBuild}}: An Automatic Approach to History-Driven Repair of Build Scripts},
}

@REPORT{tarlow2019:learning,
  AUTHOR = {Tarlow, Daniel and Moitra, Subhodeep and Rice, Andrew and Chen, Zimin and Manzagol, Pierre-Antoine and Sutton, Charles and Aftandilian, Edward},
  INSTITUTION = {CoRR e-print},
  DATE = {2019-11},
  NUMBER = {arXiv: 1911.01205},
  TITLE = {Learning to {{Fix Build Errors}} with {{Graph2Diff Neural Networks}}},
  TYPE = {techreport},
}

@REPORT{deSouza2017,
  AUTHOR = {{de Souza}, Higor A. and Chaim, Marcos L. and Kon, Fabio},
  INSTITUTION = {CoRR e-print},
  DATE = {2017-07},
  NUMBER = {arXiv: 1607.04347v2},
  TITLE = {Spectrum-Based {{Software Fault Localization}}: {{A Survey}} of {{Techniques}}, {{Advances}}, and {{Challenges}}},
  TYPE = {techreport},
}

@THESIS{Perez2018,
  AUTHOR = {Perez, Alexandre Campos},
  INSTITUTION = {University of Porto},
  DATE = {2018},
  TITLE = {Spectrum-{{Based Diagnosis}}: {{Measurements}}, {{Improvements}} and {{Applications}}},
  TYPE = {{{PhD Thesis}}},
}

@ARTICLE{heiden2019:evaluation,
  AUTHOR = {Heiden, Simon and Grunske, Lars and Kehrer, Timo and Keller, Fabian and van Hoorn, Andre and Filieri, Antonio and Lo, David},
  LANGUAGE = {en},
  DATE = {2019},
  DOI = {10.1002/spe.2703},
  ISSN = {1097-024X},
  JOURNALTITLE = {Software: Practice and Experience},
  NUMBER = {8},
  PAGES = {1197--1224},
  TITLE = {An Evaluation of Pure Spectrum-Based Fault Localization Techniques for Large-Scale Software Systems},
  VOLUME = {49},
}

@INPROCEEDINGS{Wang2015a,
  AUTHOR = {Wang, Qianqian and Parnin, Chris and Orso, Alessandro},
  LOCATION = {New York, New York, USA},
  PUBLISHER = {ACM Press},
  BOOKTITLE = {International {{Symposium}} on {{Software Testing}} and {{Analysis}} ({{ISSTA}})},
  DATE = {2015},
  DOI = {10.1145/2771783.2771797},
  ISBN = {978-1-4503-3620-8},
  PAGES = {1--11},
  TITLE = {Evaluating the Usefulness of {{IR}}-Based Fault Localization Techniques},
}

@INPROCEEDINGS{ko2008:debugging,
  AUTHOR = {Ko, Amy J. and Myers, Brad A.},
  LOCATION = {Leipzig, Germany},
  PUBLISHER = {Association for Computing Machinery},
  BOOKTITLE = {{{IEEE}}/{{ACM International Conference}} on {{Software Engineering}} ({{ICSE}})},
  DATE = {2008-05},
  DOI = {10.1145/1368088.1368130},
  ISBN = {978-1-60558-079-1},
  PAGES = {301--310},
  SHORTTITLE = {Debugging Reinvented},
  TITLE = {Debugging Reinvented: Asking and Answering Why and Why Not Questions about Program Behavior},
}

@ARTICLE{deSouza2018b,
  AUTHOR = {{de Souza}, Higor A. and Mutti, Danilo and Chaim, Marcos L. and Kon, Fabio},
  DATE = {2018},
  DOI = {10.1016/j.infsof.2017.10.014},
  ISSN = {09505849},
  JOURNALTITLE = {Information and Software Technology},
  NUMBER = {October 2016},
  PAGES = {245--261},
  TITLE = {Contextualizing Spectrum-Based Fault Localization},
  VOLUME = {94},
}

\end{document}